\documentclass[options]{JHEP3} 

\title{Noncommutative Chiral Anomaly and The Dirac-Ginsparg-Wilson Operator}

\author{Badis Ydri\\
{School of Theoretical Physics , Dublin Institute for
Advanced Studies , 10 Burlington Road , Dublin 4 , Ireland .}\\
E-mail : ydri@synge.stp.dias.ie .}

\preprint{SU-4252-747\\
\sf DIAS-02-12}

\abstract{It is  shown that the local axial anomaly in $2-$dimensions emerges
naturally if one postulates an underlying noncommutative fuzzy structure
of spacetime . In particular the Dirac-Ginsparg-Wilson relation on ${\bf S}^2_F$ is shown to contain an edge effect which corresponds precisely to the ``fuzzy'' $U(1)_A$ axial anomaly on the fuzzy sphere
. We also derive  a novel gauge-covariant expansion of the quark propagator in the form $\frac{1}{{\cal D}_{AF}}=\frac{a\hat{\Gamma}^L}{2}+\frac{1}{{\cal D}_{Aa}}$ where
$a=\frac{2}{2l+1}$ is the lattice spacing on ${\bf S}^2_F$ , $\hat{\Gamma}^L$ is the covariant noncommutative chirality and
${\cal D}_{Aa}$ is an effective Dirac operator which has essentially the same IR spectrum
as ${\cal D}_{AF}$ but differes from it on the UV modes. Most remarkably is the fact that both operators share the same limit and thus the above covariant expansion is not available in the continuum theory . The first bit in this expansion $\frac{a\hat{\Gamma}^L}{2}$ although it vanishes as it stands in the continuum limit , its contribution to the anomaly is exactly the canonical
theta term. The contribution of the propagator $\frac{1}{{\cal D}_{Aa}}$ is on the other hand equal to the toplogical Chern-Simons action which in two dimensions vanishes identically  . }


\dedicated{}

\begin{document}

\section{Introduction}

Fuzzy physics [\cite{madore,GKP} , see also \cite{ydri} and
references therein ] , like lattice gauge theories, is aiming for
a nonperturbative regularization of chiral gauge theories.
Discretization in fuzzy physics is achieved by treating the
underlying  spacetimes as phase spaces then quantizing them in a
canonical fashion which means in particular that we are
effectively replacing the underlying spacetimes by non-commutative
matrix models or fuzzy manifolds \cite{ydri}. As a
consequence , this regularization will preserve all symmetries
and topological features of the problem . Indeed a fuzzy space is
by construction a discrete lattice-like structure which serves to
regularize, it allows for an exact chiral invariance to be 
formulated  , but still the fermion-doubling problem is
completely avoided \cite{trg} .

Global chiral anomalies on these models were  , along with other
topological non-trivial field configurations, formulated in
\cite{bal,grosse,balsach,balgior} , while local anomaly on fuzzy ${\bf
S}^2_F$ was  treated first in \cite{presnajder} then in \cite{nagao1,giorgio1} . The relevance of Ginsparg-Wilson relations in noncommutative matrix models was noted first in \cite{trg} and \cite{miguel} then in \cite{nagao1} . In this article
we will show that despite the fact that the concept of point is
lacking on fuzzy ${\bf S}^2$, we can go beyond global
considerations and define a "fuzzy" axial anomaly associated with
a "fuzzy" $U(1)$ global chiral symmetry.

The plan of the paper is as follows . Section $2$ contains a
brief description of fuzzy ${\bf S}^2$ and its star product. Fuzzy $U(1)$
gauge theory and fermion action on ${\bf S}^2_F$ are introduced
in section $3$ where we also show the absence of the fermion doubling problem on ${\bf S}^2_F$. The Dirac-Ginsparg-Wilson relation on ${\bf S}^2_F$ and the corresponding fuzzy chiral transformations as well as the ``fuzzy'' $U(1)_A$ axial anomaly are discussed in section $4$. In Section $5$ we derive a novel gauge-covariant expansion of the quark propagator and show explicitly that no  analogous expansion exists in the continuum theory. The continuum limit is computed in this section where beside the canonical theta term we obtain the Chern-Simons action which vanishes identically in two dimensions. We conclude in section $6$ .

\section{The Fuzzy Sphere}
\subsection{Algebra}
The $2-$dimensional continuum sphere is the co-adjoint orbit of $SU(2)$ through the Pauli matrix ${\sigma}_3$ and thus it admits a symplectic structure which can be quantized in a canonical fashion to give the so-called fuzzy sphere. We now explain briefly this result .

We can define ${\bf S}^2$ by the projector $P=\frac{\bf 1}{2}+n_at_a$ with $t_a=\frac{{\sigma}_a}{2}$ since for example
the requirement $P^2=P$ gives precisely the  defining equation of ${\bf S}^2$ as a surface embedded in ${\bf R}^3$, namely
$\vec{n}^2=1$ . At the north pole of ${\bf S}^2$ , namely at $\vec{n}_0=(0,0,1)$, this projector is $P_0=diag(1,0)$ which projects down to
the state $|\vec{n}_0,\frac{1}{2}>=(1,0)$ of the $2-$dimensional Hilbert space ${\bf H}_{\frac{1}{2}}$ of the fundamental representation of $SU(2)$. A general point $\vec{n}$ of
${\bf S}^2$ is obtained by $\vec{n}=g\vec{n}_0$ where $g{\in}SU(2)$. The corresponding state in
${\bf H}_{\frac{1}{2}}$ is $|\vec{n},\frac{1}{2}>=g|\vec{n}_0,\frac{1}{2}>$. The projector on this state is
$P=|\vec{n},\frac{1}{2}><\vec{n},\frac{1}{2}|=gP_0g^{+}$ provided
\begin{equation}
g{\sigma}_3g^{+}=n^a{\sigma}_a.
\end{equation}
From this expression it is clear that ${\bf S}^2$ is indeed given by $
SU(2)/U(1)$ , and that points $\vec{n}{\in}{\bf S}^2$ are equivalence classes $\vec{n}=[g]=[gh]$ where $h{\in}U(1)$ , i.e ${\bf S}^2$ is the co-adjoint orbit of $SU(2)$ through ${\sigma}_3$ .

$|\vec{n},\frac{1}{2}>$ and $|\vec{n}_0,\frac{1}{2}>$ are the ``fundamental'' coherent states at $\vec{n}$ and $\vec{n}_0$ respectively. Roughly speaking these states will replace the points $\vec{n}$ and $\vec{n}_0$ when we go to the noncommutative fuzzy sphere.

It is not difficult to see that the symplectic $2$-form on ${\bf S}^2$ given by $
\omega{\equiv}-\frac{l}{2}{\epsilon}_{abc}n_cdn_a{\wedge}dn_b=ldcos\theta{\wedge}d\phi$ where $l$ is an undetremined  non-zero real number and
 with $\theta$ and $\phi$ being the usual angle coordinates,  can also be rewritten in the form  $\omega=ild\big(Tr{\sigma}_3g(\sigma,t)^{-1}dg(\sigma,t)\big)$ where $t$ is a time variable and $\sigma$ is a parameter in the range $[0,1]$ . This means in particular that the quantization of the above symplectic $2$-form $\omega$ is equivalent to the quantization of the Wess-Zumino term
\begin{eqnarray}
L=ilTr\bigg({\sigma}_3g^{-1}\dot{g}\bigg).\label{2.2}
\end{eqnarray}
Indeed if we define a triangle ${\Delta}$ in the plane $(t,\sigma)$ by its boundaries ${\partial}{\Delta}_1=(\sigma,t_1)$ , ${\partial}{\Delta}_2=(\sigma,t_2)$ and ${\partial}{\Delta}_3=(1,t)$ then it is a trivial exercise to show that
\begin{eqnarray}
S_{WZ}=\int_{\Delta}\omega=\int_{t_1}^{t_2}Ldt+il\int_{0}^1Tr{\sigma}_3\bigg(g(\sigma,t_1)^{-1}{\partial}_{\sigma}g(\sigma,t_1)-g(\sigma,t_2)^{-1}{\partial}_{\sigma}(\sigma,t_2)\bigg).
\end{eqnarray}
It is a known result that  the quantization of the above Wess-Zumino Lagrangian (\ref{2.2}) will give all $SU(2)$ irreducible representations with spins $s{\equiv}l$ , i.e the values of $l$ in the quantum theory become strictly quantized . 

The physical wave functions of the quantum system are complex valued functions on $SU(2)$ of the form
\begin{eqnarray}
{\psi}(g)=\sum_{m=-l}^lC_m<lm|U^{(l)}(g)|ll>,
\end{eqnarray}
with scalar product defined by $({\psi}_1,{\psi}_2)=\int_{SU(2)}d{\mu}(g){\psi}_1(g)^{*}{\psi}_2(g)$ where $d{\mu}$ is the Haar measure on $SU(2)$ . $U^{(l)}(g)$ is the IRR $l$ of $g{\in}SU(2)$ . Obviously $<lm|U^{(l)}(g)|ll>$ transforms as the heighest weight state $(l,l)$ under the right action $g{\longrightarrow}gg_1$ of the group $g_1{\in}SU(2)$ , while under the left action $g{\longrightarrow}g_1g$ , $\{<lm|U^{(l)}(g)|ll>\}$ transforms as a basis of the Hilbert space ${\bf H}_l$ of the $(2l+1)-$dimensional IRR of $SU(2)$. This left action is clearly generated by the usual angular momenta of $SU(2)$ in the IRR $l$ , namely
\begin{eqnarray}
[L_a,L_a]=i{\epsilon}_{abc}L_c~,~
\sum_{a=1}^3L_a^2=l(l+1).
\end{eqnarray}
Explicitly this action is given by
\begin{eqnarray}
&&[iL_a{\psi}](g)=\bigg[\frac{d}{dt}{\psi}\bigg(e^{-i\frac{{\sigma}_a}{2}t}g\bigg)\bigg]_{t=0}.
\end{eqnarray}
 The algebra ${\bf A}$ of all observables of the system is the algebra of linear operators which act on the left of ${\psi}(g)$ by left translations , i.e 
an arbitrary linear operator ${\phi}^F{\in}{\bf A}$ will admit in general an expansion of the form
\begin{eqnarray}
&&{\phi}^F=\sum_{a_1,..a_k}{\alpha}_{a_1..a_k}L_{a_1}...L_{a_k}.\label{sum}
\end{eqnarray}
The sum in (\ref{sum}) as we will check shortly is actually cut-off . The algebra ${\bf A}$ of all observables of the system is therefore a matrix algebra $Mat_{2l+1}$ , while  physical wave functions span a $(2l+1)-$dimensional Hilbert space ${\bf H}_l$ on which these observables are naturally acting [ see \cite{ydri} and references therein for more detail ] .

We define the noncommutative fuzzy sphere by Connes spectral triple $(Mat_{2l+1},{\bf H}_l,{\Delta}^F)$ \cite{connes} . The matrix algebra $Mat_{2l+1}$ is the above algebra ${\bf A}$ of $(2l+1){\times}(2l+1)$ matrices which acts on the $(2l+1)-$dimensional Hilbert space ${\bf H}_l$ of the IRR $l$ of $SU(2)$. Matrix coordinates on ${\bf S}^2_F$ are defined by \cite{madore,GKP,ydri}

\begin{eqnarray}
(n_1^{F})^2+(n_2^{F})^2+(n_3^{F})^2=1~,~
[n_a^F,n_a^F]=\frac{i}{{\sqrt{l(l+1)}}}{\epsilon}_{abc}n_c^F,
\end{eqnarray}
with
\begin{equation}
n_a^F=\frac{L_a}{\sqrt{l(l+1)}}.\label{22}
\end{equation}
A ``fuzzy'' function on ${\bf S}^2_F$ is a linear operator ${\phi}^F{\in}Mat_{2l+1}$ which can also be defined by an expansion of the form (\ref{sum}) .

Derivations on ${\bf S}^2_F$ are on the other hand defined by the generators of the adjoint
action of $SU(2)$ , in other words the derivative of the fuzzy
function ${\phi}^F$ in the space-time direction $a$ is
the commutator $[L_a,{\phi}^F]$ . This can also be put in the form
\begin{eqnarray}
AdL_a({\phi}^F)\equiv[L_a,{\phi}^F]=(L_a^L-L_a^R)({\phi}^F),
\end{eqnarray}
where $L_a^L$'s and $-L_a^R$'s are the generators of the IRR $l$ of
$SU(2)$ which act respectively on the left and on the right of the algebra
$Mat_{2l+1}$ , i.e $L_a^L{\phi}^F{\equiv}L_a{\phi}^F$ ,
$L_a^R{\phi}^F{\equiv}{\phi}^FL_a$ for any
${\phi}^F{\in}Mat_{2l+1}$. Hence the Laplacian operator ${\Delta}^F$ on the fuzzy sphere is simply given by the Casimir operator 
\begin{eqnarray}
{\Delta}^F=(L_a^L-L_a^R)^2.
\end{eqnarray}
The algebra of matrices $Mat_{2l+1}$ decomposes therefore under the action of the group $SU(2)$ as $l{\otimes}l=0{\oplus}1{\oplus}2{\oplus}..{\oplus}2l$ [ The first $l$ stands for the left action of the group while the other $l$ stands for the right action ]. As a consequence a general scalar function on ${\bf S}^2_F$ can be expanded in terms of 
polarization tensors as follows
\begin{eqnarray}
{\phi}^F=\sum_{k=0}^{2l}\sum_{m=-k}^{k}{\phi}_{km}\hat{Y}_{km}(l).\label{scalar}
\end{eqnarray}
[ For an extensive list of the properties of $\hat{Y}_{km}(l)$'s see
\cite{VKM} ] . This expansion is equivalent to (\ref{sum}) but now the cut-off is made explicit . The fact that the summation over $k$ involves only angular momenta which are ${\leq}2l$ originates from the fact that the spectrum $k(k+1)$ of the Laplacian ${\Delta}^F$ is cut-off at $k=2l$ . As we will show in this article this rotationally-invariant cut-off is non-trivial in the sense that it respects both gauge and chiral symmetries.

As one can already notice all these definitions are in very close analogy with
the case of continuum ${\bf S}^2$
where the algebra of functions ${\cal A}$ plays there the same role played here by the matrix algebra $Mat_{2l+1}$ . In fact the continuum limit
is defined by $l{\longrightarrow}{\infty}$ where the fuzzy coordinates $n_a^F$'s approach the ordinary coordinates
$n_a$'s and where the algebra $Mat_{2l+1}$ tends to the algebra ${\cal A}$ in the sense that
\begin{eqnarray}
&&{\phi}^F{\longrightarrow}\phi(\vec{n})=\sum_{k=0}^{\infty}\sum_{m=-k}^k{\phi}_{km}Y_{km}(\vec{n}).\label{5}
\end{eqnarray}
In above $Y_{km}(\vec{n}) $ stands for the canonical spherical harmonics .
Correspondingly fuzzy derivations reduce to ordinary ones on commutative ${\bf S}^2$ , i.e $adL_a({{\phi}^F}){\longrightarrow}{\cal L}_a(\phi)(\vec{n})$ , ${\cal L}_a=-i{\epsilon}_{abc}n_b{\partial}_c$ . Formally one writes ${\cal A}=Mat_{\infty}$ and think of the fuzzy sphere as having a finite number of
points equal to $2l+1$ which will diverge in the continuum limit $l{\longrightarrow}{\infty}$.

\subsection{Star Product}
To make the continuum limit more precise we will need to introduce the star product on ${\bf S}^2_F$ .
The irreducible  representation $l$ of $SU(2)$ can be obtained from the
symmetric product of $2l$ fundamental representations $\frac{1}{2}$ of $SU(2)$ . Given an element $g{\in}SU(2)$, its
$l-$representation matrix $U^{(l)}(g)$ is given as follows
\begin{equation}
U^{(l)}(g)=U^{(\frac{1}{2})}(g){\otimes}_s...{\otimes}_sU^{(\frac{1}{2})}(g),2l-times.
\end{equation}
$U^{(\frac{1}{2})}(g)$ is the spin $\frac{1}{2}$ representation of $g{\in}SU(2)$ . Clearly the states $|\vec{n}_0,\frac{1}{2}>$ and $|\vec{n},\frac{1}{2}>=g|\vec{n}_0,\frac{1}{2}>$ of ${\bf H}_{\frac{1}{2}}$ will correspond in ${\bf H}_l$ to the two states $|\vec{n}_0,l>$ and $|\vec{n},l>$
respectively such that
\begin{equation}
|\vec{n},l>=U^{(l)}(g)|\vec{n}_0,l>.\label{fundamental}
\end{equation}
To any fuzzy scalar function ${\phi}^F$ on ${\bf S}^2_F$ , i.e an operator ${\phi}^F$ acting on ${\bf H}_l$, we associate a "classical" function $<{\phi}^F>(\vec{n})$ on a classical ${\bf S}^2$  by
\begin{equation}
<{\phi}^F>(\vec{n})=<\vec{n},l|{\phi}^F|\vec{n},l>,\label{map}
\end{equation}
such that the product of two such operators ${\phi}_1^F$ and ${\phi}_2^F$ is mapped to the star product of the
corresponding two functions
\begin{equation}
<{\phi}_1^F>*<{\phi}_2^F>(\vec{n})=<\vec{n},l|{\phi}_1^F{\phi}_2^F|\vec{n},l>.\label{starproduct1}
\end{equation}
A long calculation shows that this star product is given explicitly by \cite{lee}

\begin{eqnarray}
<{\phi}_1^F>*<{\phi}_2^F>(\vec{n})&=&\sum_{k=0}^{2l}\frac{(2l-k)!}{k!(2l)!}K_{a_1b_1}....K_{a_kb_k}\frac{\partial}{{\partial}n^{a_1}}...\frac{\partial}{{\partial}n^{a_k}}<{\phi}_1^F>(\vec{n})\frac{\partial}{{\partial}n^{b_1}}...\frac{\partial}{{\partial}n^{b_k}}<{\phi}_2^F>(\vec{n})\nonumber\\
K_{ab}&=&{\delta}_{ab}-n_an_b-i{\epsilon}_{abc}n_c.\label{starproduct}
\end{eqnarray}
In these coherent states one can also compute
\begin{eqnarray}
&&<n_a^F>=\frac{1}{\sqrt{1+\frac{1}{l}}}n_a~,~<[L_a,{\phi}^F]>=({\cal L}_a<{\phi}^F>)(\vec{n}),\label{2.13}
\end{eqnarray}
and
\begin{equation}
\frac{1}{2l+1}Tr_l{\phi}_1^F{\phi}_2^F=\int_{{\bf S}^2} \frac{d{\Omega}}{4{\pi}}<{\phi}_1^F>*<{\phi}_2^F>(\vec{n}).\label{fuzzyintegral}
\end{equation}
The trace $Tr_l$ is obviously taken over the Hilbert space ${\bf H}_l$ . Remark finally that the coherent state $|\vec{n},l>$ becomes localized around the point $\vec{n}$ in the limit and as a consequence $Lim<n_a^F>=n_a$, $Lim<{\phi}^F>={\phi}$ and $Lim<[L_a,{\phi}_a]>={\cal L}_a{\phi}$ , etc . In this limit the star product reduces also to the ordinary product of functions .

\section{Fuzzy Actions}
\subsection{Gauge Fields on ${\bf S}^2_F$}
Next we would like to write down the Schwinger model on ${\bf S}^2_F$ . First we introudce $2-$dimensional gauge fields on the fuzzy sphere and their action . A vector field $\vec{A}^F$ on the fuzzy sphere can be ( similarly to scalar fields )
defined by an expansion in terms of polarization tensors of the form
\begin{eqnarray}
{A}_a^F=\sum_{k=0}^{2l}\sum_{m=-k}^{k}A_a(km)\hat{Y
}_{km}(l)~,~A_a^{F+}=A_a^F.
\label{vector}
\end{eqnarray}
In the continuum limit this expansion reduces to $A_a(\vec{n}) =\sum_{k=0}^{\infty}\sum_{m=-k}^kA_a(km)Y_{km}(\vec{n})$. Each component $A_a^F$ is a $(2l+1){\times}(2l+1)$ matrix and
the modes $A_a(km)$ are complex numbers satisfying
$A_a(km)^{*}=(-1)^mA_a(k-m)$
 where for each momentum $(km)$ the corresponding triple
$(A_1^F(km),A_2^F(km),A_3^F(km))$ transforms as an
$SO(3)-$vector. We have also to note here that a much natural expansion of vector fields on the finite dimensional fuzzy sphere can be given instead in terms of ``vector'' polarization tensors \cite{VKM}. The expansion (\ref{vector}) is however enough for the purpose of this article since we will mostly deal with the fermion action . Writing a gauge principle for this matrix vector
field is not difficult, indeed the action takes the usual form
\begin{eqnarray}
S_{YMF}&=&\frac{1}{4e^2}\frac{1}{2l+1}Tr_{l}F_{ab}^FF_{ab}^F.\label{action2}
\end{eqnarray}
The curvature $F_{ab}^F=-F_{ab}^{F+}$ is also given by the usual formula $
F_{ab}^F{\equiv}[D_a^F,D_b^F]-i{\epsilon}_{abc}D_c^F~$ with the covariant derivative $D_a^F=L_a+A_a^F$ or equivalently $
F_{ab}^F=[L_a,A_b^F]-[L_b,A_a^F]+[A_a^F,A_b^F]-i{\epsilon}_{abc}A_c^F$.
Gauge transformations are implemented by unitary transformations
acting on the $(2l+1)-$dimensional Hilbert space of the
irreducible representation $l$ of $SU(2)$. These transformations
are $D_a^F{\longrightarrow}D_a^{F'}=U^FD_a^FU^{F+}$ ,
$A_a^F{\longrightarrow}A_a^{F'}=U^FA_a^FU^{F+}+U^F[L_a,U^{F+}]$
and $ F_{ab}^F{\longrightarrow}F_{ab}^{F'}=U^FF_{ab}^FU^{F+}$
where $U^F=e^{i{\Omega}^F}$ and ${\Omega}^F={\Omega}^{F+}$ is
an element of the algebra $Mat_{2l+1}$ of $(2l+1){\times}(2l+1)$
matrices . $U^F$'s define then fuzzy $U(1)$
gauge theory . Clearly $U(1)_F{\equiv}U(2l+1)$ .

It is not difficult to convince ourselves that  (\ref{action2}) has the correct continuum limit , namely
\begin{eqnarray}
S_{YMF}{\longrightarrow}S_{YM}=\frac{1}{4e^2}\int_{{\bf S}^2} \frac{d{\Omega}}{4{\pi}}F_{ab}F_{ab}~,~l{\longrightarrow}{\infty}.
\end{eqnarray}
For example one can use the star product (\ref{starproduct}) on ${\bf S}^2_F$ to see that $A_a=Lim_{l{\longrightarrow}{\infty}}<A_a^F>$ and $F_{ab}=Lim_{l{\longrightarrow}{\infty}}<F_{ab}^F>={\cal L}_aA_b-{\cal L}_bA_a-i{\epsilon}_{abc}A_c$ . Also by using (\ref{fuzzyintegral}) it is seen that the trace $\frac{1}{2l+1}Tr_l$ behaves as the integral $\int_{{\bf S}^2}\frac{d{\Omega}}{4{\pi}}$ in the limit where the star product becomes the commutative product .

A final remark concerning vector fields is to note that the gauge field $\vec{A}^F$ has
three components and hence an extra condition is needed in order
to project this gauge field onto two dimensions . One adopts here
the prescription of \cite{nair}, i.e we impose on the gauge
filed $\vec{A}^F$ the gauge-covariant condition
\begin{eqnarray}
D_a^FD_a^F=l(l+1).\label{local}
\end{eqnarray}
This constraint reads explicitely $\{n_a^F,A_a^F\}=-(\vec{A}^F)^2/\sqrt{l(l+1)}$ , and thus it is not difficult to check that in the continuum limit $l{\longrightarrow}{\infty}$ the normal component of the gauge field is zero , i.e ${\phi}{\equiv}\vec{n}.\vec{A}=0$ .

\subsection{Fermion Doubling}

In close
analogy with the free fermion action on ordinary ${\bf S}^2$
, free fermion action on fuzzy ${\bf S}^2$ is defined by
\begin{eqnarray}
S_F&=&\frac{1}{2l+1}Tr_{l}\bar{{\psi}}_F{\cal D}_F{\psi}_F~,~ {\cal D}_{
F}={\sigma}_a[L_a,...]+1.\label{fuzziii}
\end{eqnarray}
${\cal D}_F$ is precisely the Dirac operator on
fuzzy ${\bf
 S}^2$
\cite{ydri,trg,bal,grosse} , and ${\sigma}_i$'s are Pauli matrices
 . The fuzzy
spinor ${\psi}_F$ is an element of $Mat_{2l+1}{\otimes}{\bf C}^2$ , it is
of mass
dimension $({\rm mass})^{\frac{1}{2}}$ and is such that
$\bar{\psi}_F={\psi}_F^{+}$
.

The so-called
Grosse-Klim\v{c}\'{i}k-Pre\v{s}najder Dirac operator ${\cal D}_F$ on
fuzzy ${\bf S}^2$ admits a chirality operator which can be
seen as follows , first we rewrite ${\cal D}_F$ in the form
\cite{trg,ydri, bal,grosse}
\begin{eqnarray}
\frac{1}{2l+1}{\cal D}_{F}&=&\frac{1}{2}({\Gamma}^R+{\Gamma}^L),\label{key}
\end{eqnarray}
where ${\Gamma}^R$ and ${\Gamma}^L$
are the operators
\begin{eqnarray}
{\Gamma}^L&=&\frac{1}{l+\frac{1}{2}}[\vec{\sigma}.\vec{L}^L+\frac{1}{2}]~,~
{\Gamma}^R=\frac{1}{l+\frac{1}{2}}[-\vec{\sigma}.\vec{L}^R+\frac{1}{2}].\label{chiralities}
\end{eqnarray}
By analogy with (\ref{22}) we also
define $n_a^L=\frac{L_a^L}{\sqrt{l(l+1)}}$ and
$n_a^R=\frac{L_a^R}{\sqrt{l(l+1)}}$ with obvious continuum limits
, i.e $n_a^L,n_a^R{\longrightarrow}n_a$ when
$l{\longrightarrow}{\infty}$ ( since the left and right actions
become identical in the limit). Remark that one can also make the
identification $L_a^L{\equiv}L_a$ , $n_a^L{\equiv}n_a^F$. In
general all operators acting from the left can be thought of as
elements of the algebra $Mat_{2l+1}$.

${\Gamma}^R$ is the chirality operator as was shown originally in
\cite{watamuras} , this choice is also motivated by the fact that
${\Gamma}^R{\longrightarrow}-{\gamma}=-\vec{\sigma}\vec{n}~,~{\rm
when}~l{\longrightarrow}{\infty}$ ,
$({\Gamma}^R)^2={\Gamma}^{R}~,~({\Gamma}^{R})^{+}={\Gamma}^{R}~,~{\rm
and}~[{\Gamma}^R,{\phi}^F]=0$ for any ${\phi}^F{\in}Mat_{2l+1}$ .
However this ${\Gamma}^R$ does not exactly anticommute with the
Dirac operator since
\begin{eqnarray}
{\cal D}_F{\Gamma}^R+{\Gamma}^R{\cal D}_{F}&=&\frac{1}{l+\frac{1}{2}}{\cal D}^2_{F}.\label{GWrelation}
\end{eqnarray}
The continuum limit of this equation is simply given by the canonical anticommutation relation $\{{\cal D},{\gamma}\}=0$ where the continuum Dirac operator ${\cal D}$ is given by ${\cal D}={\sigma}.{\cal L}+1$. Remark that for all practical purposes ${\Gamma}^L$ is also a chirality operator , it has the correct continuum limit , i.e ${\Gamma}^L{\longrightarrow}{\gamma}$ when $l{\longrightarrow}{\infty}$ , it satisfies $({\Gamma}^{L})^2=1$ , $({\Gamma}^{L})^{+}={\Gamma}^L$ , and by using (\ref{key}) one rewrites (\ref{GWrelation}) in the form
\begin{eqnarray}
{\cal D}_F{\Gamma}^L+{\Gamma}^L{\cal D}_{F}&=&\frac{1}{l+\frac{1}{2}}{\cal D}^2_{F}.\label{GWrelation2}
\end{eqnarray}
Indeed ${\Gamma}^L$ fails only to commute with the elements of the algebra $Mat_{2l+1}$ . For later use we notice that (\ref{GWrelation}) and (\ref{GWrelation2}) can also be rewritten in the form
\begin{eqnarray}
{\cal D}_F{\Gamma}^R-{\Gamma}^L{\cal D}_F=0.\label{GWrelation3}
\end{eqnarray}
It was shown in \cite{ydri,trg} that the pair $({\Gamma}^R,{\Gamma}^L)$
defines a chiral structure on fuzzy ${\bf S}^2$ which satisfies
$a)$ the Ginsparg-Wilson relation , $b)$ is without fermion
doubling and $c)$ has the correct continuum limit
. Indeed ${\Gamma}^R$ and ${\Gamma}^L$ together with the identity generate a Ginsparg-Wilson algebra  where the canonical Dirac-Ginsparg-Wilson  operator is defined by ${\cal D}_{DGW}={\Gamma}^R{\cal D}_F$ while the lattice spacing is identified as $a=\frac{2}{2l+1}$ [ see \cite{balgior} and references therein for more detail ] . The Ginsparg-Wilson relation for zero gauge field is essentially given by equation (\ref{GWrelation3}) .

On the other hand the absence of fermion doubling  can
be seen by comparing the spectrum of ${\cal D}_{F}$ which can be easily computed to be given by
\begin{eqnarray}
{\cal D}_{F}(j)=\{{\pm}(j+\frac{1}{2}) ,
j=\frac{1}{2},\frac{3}{2},...,2l-\frac{1}{2}\}\cup
\{j+\frac{1}{2}~~,~~~j=2l+\frac{1}{2}\}
\end{eqnarray}
 with the spectrum of the continuum Dirac operator ${\cal D}$  given by ${\cal D}(j)=\{{\pm}(j+\frac{1}{2}) ,
j=\frac{1}{2},\frac{3}{2},...,\infty \}$ \cite{grosse} . As one
can immediately notice there is no fermion doubling and the spectrum
of ${\cal D}_F$ is simply cut-off at the top eigenvalue
$j=2l+\frac{1}{2}$. The continuum limit of this chiral structure is therefore obvious by construction .

The fuzzy gauged Dirac operator is obviously defined by ${\cal D}_{AF}={\cal D}_{F}+{\sigma}_aA_a^F$ and thus the fuzzy gauged action  is
given by
\begin{eqnarray}
S_{AF}&=&\frac{1}{2l+1}Tr_l\bar{\psi}_F{\cal D}_{AF}{\psi}_F{\equiv}\frac{1}{2l+1}Tr_{l}\bigg[\bar{{\psi}}_F{\sigma}_{a}[L_a,{\psi}_F]+\bar{\psi}_F{\psi}_F+\bar{{\psi}}_F{\sigma}_a{A}_a^F{{\psi}}_F\bigg].\label{gaugeFaction}\nonumber\\
\end{eqnarray}
The spinor ${\psi}_F$ is assumed here to
transform in the fundamental representation of the fuzzy gauge group $U(1)_F{\equiv}U(2l+1)$ , i.e
${\psi}_F{\longrightarrow}{\psi}^{'}_{F}=U^F{\psi}_{F}$,
$\bar{\psi}_F{\longrightarrow}
\bar{\psi}^{'}_{F}=\bar{\psi}_{F}U^{F+}$.

Again it is not difficult to see that both classical actions (\ref{fuzziii}) and (\ref{gaugeFaction}) behave correctly in the continuum limit in the sense that $S_F{\longrightarrow}S=\int_{{\bf S}^2} \frac{d{\Omega}}{4{\pi}}\bar{\psi}{\cal D}{\psi}$ and $S_{AF}{\longrightarrow}S_A=\int_{{\bf S}^2} \frac{d{\Omega}}{4{\pi}}\bar{\psi}{\cal D}_A{\psi}$ when $l{\longrightarrow}{\infty}$ and where ${\psi}=Lim_{l{\longrightarrow}{\infty}}<{\psi}^F>$ and ${\cal D}_A={\cal D}+{\sigma}_aA_a$ . Explicitly we write
\begin{eqnarray}
S=\int_{{\bf S}^2} \frac{d{\Omega}}{4{\pi}} \bigg[\bar{\psi}{\sigma}_a{\cal
L}_a({\psi})+\bar{\psi}{\psi}\bigg],\label{classiii}
\end{eqnarray}
and
\begin{eqnarray}
S_{A}&=&\int
\frac{d{\Omega}}{4{\pi}}\bigg[\bar{\psi}{\sigma}_a{\cal
L}_a({\psi})+\bar{\psi}{\psi}+\bar{\psi}{\sigma}_aA_a{\psi}\bigg].\label{contaction1}
\end{eqnarray}
By putting the actions (\ref{action2}) and (\ref{gaugeFaction}) together we obtain the fuzzy Schwinger model on ${\bf S}^2_F$ , namely
\begin{eqnarray}
S_{Schwinger}=\frac{1}{4e^2}\frac{1}{2l+1}Tr_{l}F_{ab}^FF_{ab}^F+\frac{1}{2l+1}Tr_l\bar{\psi}_F{\cal D}_{AF}{\psi}_F,
\end{eqnarray}
where the gauge field $\vec{A}^F$ is also assumed to satisfy the constraint (\ref{local}) .

\section{Quantum Chiral Symmetry on ${\bf S}^2_F$}
\subsection{Fermion Propagator}
The quantum
theory of interest is defined through the following path integral
\begin{eqnarray}
\int{\cal D}A_i^F\int {\cal D}{{\psi}_{F}}{\cal
D}\bar{{\psi}}_{F}e^{-S_{Schwinger}}.\label{par}
\end{eqnarray}
The anomaly arises generally from the
non-invariance of the fermionic measure under chiral transformations
\cite{fujikawa} and hence we will focus first on  this measure and show
explicitly that for all finite approximations of the
noncommutative Schwinger model on ${\bf S}^2_F$ this measure is in fact exactly invariant . Indeed we will show shortly that the $U(1)$ ``fuzzy''axial anomaly on ${\bf S}^2_F$ comes entirely from the non-invariance of the action due to edge effect . In the appendix we will also discuss how one can shift the anomaly from the action back to the measure.

It is also enough to treat in the following the gauge field as a background field since all we want to compute is the fuzzy axial anomaly on ${\bf S}^2_F$  and its continuum limit the canonical local axial anomaly on ${\bf S}^2$ , i.e only the fermion loop is relevant . Quantization of the Yang-Mills action $S_{YMF}$ will be reported elsewhere .

In a matrix model such as (\ref{par}) manipulations on
the quantum measure have a precise meaning . Indeed and by
following \cite{fujikawa} we first expand the fuzzy spinors
${\psi}_{F}$ , $\bar{\psi}_{F}$ in terms of the eigentensors
${\phi}(\mu,A)$ of the Dirac operator ${\cal D}_{AF}$, write $
{\psi}_{F}=\sum_{\mu}{\theta}_{\mu}{\phi}(\mu,A)$ , $
\bar{\psi}_{F}=\sum_{\mu}\bar{\theta}_{\mu}{\phi}^{+}(\mu,A)$
where ${\theta}_{\mu}$'s , $\bar{\theta}_{\mu}$'s are independent
sets of Grassmanian variables, and ${\phi}(\mu,A)$'s are defined
by ${\cal D}_{AF}{\phi}(\mu,A)={\lambda}_{\mu}(A){\phi}(\mu,A)$ ,
 and
normalized such that
\begin{eqnarray}
\frac{1}{2l+1}Tr_{l}{\phi}^{+}(\mu,A){\phi}(\nu,A)={\delta}_{\mu
\nu } .\label{nst}
\end{eqnarray}
 For zero fuzzy gauge fields $\mu$ stands for $j$ , $k$
and $m$ which are the eigenvalues of
$\vec{J}^2=(\vec{K}+\frac{\vec{\sigma}}{2})^2$ ,
$\vec{K}^2=(\vec{L}^L-\vec{L}^R)^2$ and $J_3$ respectively . Indeed the asymptotic behaviour when
$A^F_a{\longrightarrow}0$ of ${\phi}(\mu,A)$'s and
${\lambda}_{\mu}(A)$'s is given by
${\lambda}_{\mu}(A){\longrightarrow}j(j+1)-k(k+1)+\frac{1}{4}$ and
$
{\phi}(\mu,A){\longrightarrow}\sqrt{2l+1}\sum_{k_3,\sigma}C^{jm}_{kk_3\frac{1}{2}\sigma}\hat{Y}_{kk_3}(l)
{\chi}_{\frac{1}{2}\sigma}$ \cite{ydri,VKM,grosse} . The quantum
measure is therefore well defined  and it is given by
\begin{eqnarray}
&&{\cal D}{\psi}_{F}{\cal
D}\bar{\psi}_{F}=\prod_{\mu}d{\theta}_{\mu}d{\bar{\theta}_{\mu}}{\longrightarrow}\prod_{k=0}^{2l}\prod_{j=k-\frac{1}{2}}^{k+\frac{1}{2}}\prod_{m=-j}^{j}d{\theta}_{kjm}d{\bar{\theta}_{kjm}}
\end{eqnarray}
From the action $S_{AF}$ and the identity (\ref{nst}) one can compute the propagator $<{\theta}_{\mu}\bar{\theta}_{\nu}>={\delta}_{\mu \nu}/{\lambda}_{\mu}(A)$ or equivalently
\begin{eqnarray}
<{\psi}_{F\alpha}^{AB}\bar{\psi}_{F\beta}^{CD}>_{ev}&=&\sum_{\mu}\frac{1}{{\lambda}_{\mu}(A)}{\phi}_{\alpha}^{AB}(\mu,A){\phi}_{\beta}^{+CD}(\mu,A){\equiv}(2l+1)\bigg(\frac{1}{
{\cal D}_{AF}}\bigg)_{\alpha \beta}^{AB,DC}.\label{prop}
\end{eqnarray}
$<>_{ev}$ stands for expectation value . We are assuming no monopole configurations and thus the inverse of the gauged Dirac operator is always well defined .  This inverse is defined by the formula
\begin{eqnarray}
({\cal D}_{AF})_{\gamma \alpha}^{C^{'}D^{'},AB}(\frac{1}{{\cal D}_{AF}})_{\alpha \beta}^{AB,DC}={\delta}_{\gamma \beta}{\delta}^{C^{'}D}{\delta}^{D^{'}C}.
\end{eqnarray}
In above we have also used the fact that because the Dirac operator ${\cal D}_{AF}$ is self-adjoint on
$Mat_{2l+1}{\otimes}{\bf C}^2$ , the states ${\phi}(\mu,A)$'s must form
a complete set , viz
\begin{eqnarray}
\frac{1}{2l+1}\sum_{\mu}{\phi}^{AB}_{\alpha}(\mu,A){\phi}^{+CD}_{\beta}(\mu,A)={\delta}_{\alpha
\beta}{\delta}^{AD}{\delta}^{BC}.\label{complete}
\end{eqnarray}
The Dirac operator $({\cal D}_{AF})_{\alpha \beta}$ and the propagator $({\cal D}_{AF}^{-1})_{\alpha \beta}$ carry $4$ indices because they can act on matrices of $Mat_{2l+1}$ either from the left or from the right , for example ${\cal D}_{AF}$ acts explicitly as $({\cal D}_{AF}{\phi}(A,\mu))^{AB}_{\alpha}=({\cal D}_{AF})_{\alpha \beta}^{AB,CD}{\phi}_{\beta}^{CD}$ where
\begin{eqnarray}
({\cal D}_{AF})_{\alpha \beta}^{AB,CD}=({\sigma}_a)_{\alpha \beta}(D_a^F)^{AC}{\delta}^{BD}-({\sigma}_a)_{\alpha \beta}(L_a)^{DB}{\delta}^{AC}+{\delta}_{\alpha \beta}{\delta}^{AC}{\delta}^{BD}.
\end{eqnarray}
\subsection{The Dirac-Ginsparg-Wilson Relation on  ${\bf S}^2_F$}

We will now undertake the task of deriving the Dirac-Ginsparg-Wilson relation on ${\bf S}^2_F$ in the presence of a gauge field . As it turns out this relation contains exactly a fuzzy anomaly which will become a local anomaly in the limit . We first start with the free theory
and rewrite the Ginsparg-Wison relation (\ref{GWrelation3}) in the form $
({\Gamma}^R-{\Gamma}^L){\cal D}_F+{\cal D}_F({\Gamma}^R-{\Gamma}^L)=0$.This
means that in the absence of gauge fields we must have $
tr[{\Gamma}^R-{\Gamma}^L]=0$  where the trace is taken in the space
of spinors  , in other words the anomaly vanishes.
However if we include the gauge field through the gauge-covariant
Dirac operator ${\cal D}_{AF}$ which can also be written in the form
\begin{eqnarray}
\frac{1}{2l+1}{\cal D}_{AF}=\frac{1}{2}({\Gamma}^R+\hat{\Gamma}^L),\label{comp}
\end{eqnarray}
with
\begin{eqnarray}
\hat{\Gamma}^L=\frac{1}{l+\frac{1}{2}}\bigg[{\sigma}_aD_a^F+\frac{1}{2}\bigg]={\Gamma}^L+\frac{1}{l+\frac{1}{2}}\vec{\sigma}.\vec{A}^F.\label{like}
\end{eqnarray}
[ (\ref{comp}) is to be compared with the free formula (\ref{key})] .
Then we can
compute instead the gauge-covariant anticommutation relation
\begin{eqnarray}
&&\{{\Gamma}^R-\hat{\Gamma}^L,{\cal D}_{AF}\}=-\frac{4}{2l+1}\bigg(F^F+(D_a^F)^2-l(l+1)\bigg)\nonumber\\
&&F^F=\frac{i}{2}{\epsilon}_{abc}{\sigma}_cF_{ab}^F.\label{ginsparg-wilson}
\end{eqnarray}
The continuum limit of this equation is $\{{\gamma},{\cal
D}_A\}=2{\phi}=0$ where $\phi$ is the normal component of the gauge field on ${\bf S}^2$ which is zero by the continuum limit of (\ref{local}). In other words , in the continuum interacting
theory one might be tempted to conclude that $tr{\gamma}=0$ which
we know is wrong in the presence of gauge fields. Noncommutative
geometry , as it is already obvious from equation
(\ref{ginsparg-wilson}), already gives us the structure of  the
chiral anomaly, indeed by using the constraint (\ref{local}) one
can put (\ref{ginsparg-wilson}) in the equivalent form
\begin{eqnarray}
{\cal D}_{AF}{\Gamma}^R-\hat{\Gamma}^L{\cal D}_{AF}=-\frac{2}{2l+1}F^F=-\frac{i}{2l+1}{\epsilon}_{abc}{\sigma}_cF_{ab}^F.\label{dicta}
\end{eqnarray}
We recognize immediately the left hand side as a fuzzy anomaly
since it vanishes in the limit $A^F{\longrightarrow}0$ where
(\ref{dicta}) reduces to (\ref{GWrelation3}) . We will now show
that this corresponds indeed to the actual global $U_A(1)$ fuzzy
anomaly on ${\bf S}^2_F$ .

In the following we will also need to write down the explicit
action of the operators ${\Gamma}^{R}$ , $\hat{\Gamma}^L$ . We
have
\begin{eqnarray}
&&({\Gamma}^R)_{\alpha \beta}^{AB,CD}=\frac{{\delta}^{AC}}{l+\frac{1}{2}}\bigg(-({\sigma}_a)_{\alpha \beta}L_a^{DB}+\frac{1}{2}{\delta}_{\alpha \beta}{\delta}^{DB}\bigg)\nonumber\\
&&(\hat{\Gamma}^L)_{\alpha
\beta}^{AB,CD}=\frac{{\delta}^{BD}}{l+\frac{1}{2}}\bigg(({\sigma}_a)_{\alpha
\beta}(D_a^F)^{AC}+\frac{1}{2}{\delta}_{\alpha
\beta}{\delta}^{AC}\bigg){\equiv}{\delta}^{BD}(\hat{\Gamma}^L)_{\alpha
\beta}^{AC}.\label{used}
\end{eqnarray}
In particular from the second equation we note that the operator
$(\hat{\Gamma}^L)_{\alpha \beta}$ since it acts from the left it
can also be thought of as an element of the algebra $Mat_{2l+1} $.

\subsection{Gauge-Covariant Axial Current }
Now we need to define chiral symmetry on the fuzzy sphere which must also  be consistent with gauge invariance. In the absence of gauge fields and motivated by (\ref{GWrelation3}) we define chiral transformations by
\begin{eqnarray}
&&{\psi}_F{\longrightarrow}{\psi}^{'}_F={\psi}_F+({\Gamma}^{R}{\lambda}^F{\psi}_F)+O(({\lambda}^F)^2)~,
~\bar{\psi}_F{\longrightarrow}\bar{\psi}^{'}_F=\bar{\psi}_F-(\bar{\psi}_F{\lambda}^F{\Gamma}^L)+O(({\lambda}^F)^2).\label{fuzzytrans20}\nonumber\\
\end{eqnarray}
It is then not difficult to check that the action $S_{F}$ changes
by a total divergence. The chiral parameter
${\lambda}^F$ is an arbitrary matrix in $Mat_{2l+1}$ . In the
presence of gauge fields it is therefore obvious that the minimal
prescription is given by
\begin{eqnarray}
&&{\psi}_F{\longrightarrow}{\psi}^{'}_F={\psi}_F+({\Gamma}^{R}{\lambda}^F{\psi}_F)+O(({\lambda}^F)^2)~,~
\bar{\psi}_F{\longrightarrow}\bar{\psi}^{'}_F=\bar{\psi}_F-(\bar{\psi}_F{\lambda}^F\hat{\Gamma}^L)+O(({\lambda}^F)^2),\label{fuzzytrans2}\nonumber\\
\end{eqnarray}
where $\hat{\Gamma}^L$ is the $A-$dependent  chirality-like
operator defined in (\ref{like}) . These transformations are of
course motivated by (\ref{dicta}) . Indeed one can check that
(\ref{fuzzytrans2}) reduces in the limit to the usual chiral
transformations yet it guarantees gauge invariance in the
noncommutative fuzzy setting since both $\hat{\Gamma}^L$ and the
chiral parameter ${\lambda}^F$ transform covariantly as
$U^F\hat{\Gamma}^LU^{F+}$ and $U^F{\lambda}^FU^{F+}$ respectively
under gauge transformations . Naive fuzzy chiral transformations
which would be again given by (\ref{fuzzytrans2}) but with
${\Gamma}^L$ instead of $\hat{\Gamma}^L$ are in fact inconsistent
with gauge symmetry as one might easily convince ourselves ,
whereas in the case of (\ref{fuzzytrans2}) if one gauge transform
${\psi}_F$ and $\bar{\psi}_F$ by a unitary transformation $U^{F}$
their chiral transform ${\psi}_F^{'}$ and $\bar{\psi}^{'}_F$ will
also rotate by the same gauge transformation $U^{F}$. For
completeness we write the meaning of (\ref{fuzzytrans2})
explicitly as follows
\begin{eqnarray}
&&{\psi}^{'}_{F\alpha}={\psi}_{F\alpha}+\frac{1}{2l+1}\bigg(-2({\sigma}_a)_{\alpha \beta}{\lambda}^F{\psi}_{F\beta}L_a+{\lambda}^F{\psi}_{F\alpha}\bigg)+O(({\lambda}^F)^2)\nonumber\\
&&\bar{\psi}^{'}_{F\alpha}=\bar{\psi}_{F\alpha}-\frac{1}{2l+1}\bigg(2\bar{\psi}_{F\beta}{\lambda}^F({\sigma}_a)_{\beta
\alpha}D_a^F+\bar{\psi}_{F\alpha}{\lambda}^F\bigg)+O(({\lambda}^F)^2).
\end{eqnarray}
The  change of the action under these fuzzy chiral transformations (\ref{fuzzytrans2}) is given by
\begin{eqnarray}
\int {\cal D}{{\psi}_{F}}^{'}{\cal
D}\bar{{\psi}}_{F}^{'}e^{-S^{'}_{AF}}=\int
{\cal D}{{\psi}_{F}}{\cal
D}\bar{{\psi}_{F}}e^{{S}_{{\theta}F}}e^{-S_{AF}-{\Delta}S_{AF}},\label{partition0}
\end{eqnarray}
with
\begin{eqnarray}
{\Delta}S_{AF}&=&-\frac{1}{2l+1}Tr_{l}{\lambda}^F[D_a^F,{\cal J}_a^5]-\frac{i}{(2l+1)^2}{\epsilon}_{abc}Tr_l\bar{\psi}_F{\lambda}^F{\sigma}_cF_{ab}^F{\psi}_F\nonumber\\
{\cal J}_a^5&=&\bar{\psi}_F{\sigma}_a{\Gamma}^R{\psi}_F-[(\bar{\psi}_F{\sigma}_a)_{\alpha},({\Gamma}^R{\psi}_F)_{\alpha}].
\label{divergence}
\end{eqnarray}
An immediate remark is that the change in the action is not simply a total covariant divergence but there is an extra piece which vanishes ( as it stands ) only in the continuum limit, i.e we have a gauge-invariant edge effect. It is clear that the source of this edge effect is the RHS of the Ginsparg-Wilson relation (\ref{dicta}).

The theta term in the path integral (\ref{partition0}) is also gauge-invariant and it is given explicitly  by
\begin{eqnarray}
S_{{\theta}F}&=&-\frac{1}{2l+1}Tr_{l}\sum_{\mu}{\phi}^{+}(\mu,A){\lambda}^F({\Gamma}^R-\hat{\Gamma}^L){\phi}(\mu,A).\label{anomaly}
\end{eqnarray}
[This is gauge-invariant since  for example ${\phi}(\mu , A)$ must transform as $U^L{\phi}(\mu,A)$ in order for the Dirac equation to be gauge-invariant]. Due to the finiteness of the matrix model , it is an identity
easy to check   that the theta term ${S}_{\theta F}$ is zero ,
indeed by using (\ref{complete}) and (\ref{used}) we compute
\begin{eqnarray}
S_{\theta F}=2Tr_l\bigg({\lambda}^Ftr_2(\sigma
D^F)\bigg)+\frac{2}{2l+1}Tr_l\big({\lambda}^F\big)Tr_l\bigg(tr_2({\sigma}L)\bigg){\equiv}0.\label{ref}
\end{eqnarray}
$tr_2$ is the $2-$dimensional spin trace , i.e $tr_2{\bf 1}=2$ ,
$tr_2{\sigma}_a=0$ , etc . In fact it is this trace which actually
vanishes, and hence it seems that the chiral WT identity ${\Delta}S_{GF}=S_{\theta F}$ is not anomalous due to the non-invariance of the measure under chiral transformations but rather anomalous due to edge effects ( the second term of the first equation of (\ref{divergence}) ) . We write this WT identity ${\Delta}S_{GF}=0$ in the form
\begin{eqnarray}
<[D_a^F,{\cal J}_a^5]>_{ev}=\frac{i}{2l+1}{\epsilon}_{abc}<({\sigma}_c)_{\alpha \beta}F_{ab}^F{\psi}_{F\beta}\bar{\psi}_{F\alpha}>_{ev}.
\end{eqnarray}
Using the propagator (\ref{prop}) we find the anomaly
\begin{eqnarray}
[D_a^F,<{\cal J}_a^5>]^{CB}=i{\epsilon}_{abc}(F_{ab}^F)^{CD}tr_2{\sigma}_c(D_{AF}^{-1})^{DA,BA}.\label{ep}
\end{eqnarray}
This is the ``fuzzy'' form of the global $U_A(1)$ axial anomaly on ${\bf S}^2_F$ . The inetgrated form of this anomaly is clearly given by

\begin{eqnarray}
{\cal A}_{{\theta}F}
&=&-\frac{i}{2l+1}{\epsilon}_{abc}({\lambda}^F)^{BC}(F_{ab}^F)^{CD}tr_2{\sigma}_c({\cal
D}_{AF}^{-1})^{DA,BA}.\label{alt}
\end{eqnarray}
The central claim of this article is that higher modes of the fuzzy sphere are essentially the source of the anomaly. In fact these top modes are the source of the edge effects we saw in (\ref{dicta}) and in (\ref{divergence}) which effectively yielded the non-vanishing anomaly (\ref{ep}) . Remark that if we make the top modes larger , i.e $l{\longrightarrow}{\infty}$ , these effects in (\ref{dicta}) and (\ref{divergence}) become smaller while the fuzzy anomaly (\ref{ep}) remains non-zero . In the strict limit $l{\longrightarrow}{\infty}$ these edge effects in (\ref{dicta}) and (\ref{divergence}) completely disappear while the anomaly (\ref{ep}) survives [ see below for the explicit proof ]. This is the origin of the anomaly in this context .

\section{The Continuum Limit}
\subsection{Gauge Covariant Expansion on ${\bf S}^2_F$}
By definition the local axial anomaly on ${\bf S}^2$ is the continuum limit $l{\longrightarrow}{\infty}$ of the fuzzy axial anomaly (\ref{ep}) given also in (\ref{alt}).
From equation (\ref{alt}) one can immediately notice that the computation ( perturbative or otherwise ) of the exact Dirac propagator $({\cal D}_{AF})^{-1}$ is needed in order to obtain a closed formula of the fuzzy anomaly and its continuum limit. Towards this end the best approach is as usual to expand the above propagator in a gauge covariant manner and then calculate the anomaly. As it turns out there exists a gauge covariant expansion on the noncommutative fuzzy sphere which is not available  in the continuum setting and which yields a non-perturbative result in the limit . In fact it is precisely in this sense that the fuzzy sphere is said to be a gauge-invariant, chiral-invariant regularization of the continuum physics . This is also in contrast with other approximation schemes in which gauge covariant expansions are so often absent .

This  gauge covariant expansion can be motivated as follows . Starting with zero gauge field one can see that the free Dirac propagator $\frac{1}{{\cal D}_{F}}$ admits the expansion
\begin{eqnarray}
\frac{1}{{\cal D}_F}&=&\frac{1}{a}\frac{1}{{\cal D}_F^2}({\Gamma}^R+{\Gamma}^L)=\frac{a{\Gamma}^L}{2}+\frac{1}{{\cal D}_{a}},\label{result0}
\end{eqnarray}
with
\begin{eqnarray}
~\frac{1}{{\cal D}_{a}}=ib\frac{1}{|{\cal D}_F|^2}{\cal D}_{F}^{'}{\Gamma}^L.\label{result00}
\end{eqnarray}
In above we  have used the result
${\Gamma}^R{\Gamma}^{L}=-1+\frac{a^2}{2}\big({\cal D}^2_F+2i\frac{b}{a}{\cal D}_{F}^{'}\big)$ as well as equation (\ref{GWrelation3}) where $a$ is the lattice spacing introduced before , i.e $a=\frac{2}{2l+1}$ , and $b=\sqrt{1-\frac{a^2}{4}}$.

${\cal D}_{F}^{'}$ is the
Watamura Dirac operator given by $
{\cal D}_{F}^{'}={\epsilon}_{abc}{\sigma}_an_b^FL_c^R$ \cite{ydri,watamuras}.
The continuum limit ${\cal D}^{'}$ of ${\cal D}_F^{'}$ is related to
${\cal D}$ ( the continuum limit of ${\cal D}_F$ ) by ${\cal D}^{'}=i{\gamma}{\cal D}$ and hence both
operators ${\cal D}^{'}$ and ${\cal D}$ have the same spectrum . This
does not mean that they commute since in fact we have $\{{\cal
D}^{'},{\cal D}\}=0$  . In the fuzzy , the spectrum of $({\cal D}_{F})^2$ as we have seen is simply cut-off at the top
modes $j=2l+\frac{1}{2}$ and is given by $(j+\frac{1}{2})^2$ while the spectrum of $({\cal D}_F^{'})^2$ is deformed given by
\begin{eqnarray}
({\cal D}_F^{'}(j))^2=\bigg\{(j+\frac{1}{2})^2\big[1+\frac{1-(j+\frac{1}{2})^2}{4l(l+1)}\big]~,~j=\frac{1}{2},\frac{3}{2},...,2l+\frac{1}{2}\bigg\}.
\end{eqnarray}
In
particular the eigenvalues of ${\cal D}_F^{'}$ when $j=2l+\frac{1}{2}$ are
now exactly zero  while for other large $j$'s these eigenvalues
are very small \cite{watamuras}. As a consequence
of this behaviour we have in fact the exact anticommutation relation
$\{{\Gamma}^R,{\cal D}_{F}^{'}\}=0$ .

Next we remark that the ``Dirac'' operator ${\cal D}_{a}$ defined in (\ref{result00}) is such that ${\cal D}_{a}^{-2}={\cal D}_F^{-2}-\frac{a^2}{4}$ . In other words on the top modes $j=2l+\frac{1}{2}$ , $({\cal D}_{a})^2$ is strictly infinite whereas for the other large $j$'s the eigenvalues of $({\cal D}_{a})^2$  are quite large. On the IR modes the spectrum of $({\cal D}_{a})^2$ is essentially equal to that of ${\cal D}_F^2$ . Indeed one can  explicitly compute the spectrum of  $({\cal D}_{a})^2$ and one finds the result
\begin{eqnarray}
({\cal D}_a(j))^2=\bigg\{(j+\frac{1}{2})^2\frac{(2l+1)^2}{(2l+\frac{1}{2}-j)(2l+\frac{3}{2}+j)}~,~j=\frac{1}{2},\frac{3}{2},...,2l+\frac{1}{2}\bigg\}.
\end{eqnarray}
This means in particular that as far as the second term in (\ref{result0}) is concerned modes with large $j$'s
do not effectively propagate . We will show ( after we also take the gauge field into account ) that the contribution
of these  UV modes to the anomaly  translates in the continuum limit into the contribution of contact terms whereas the
 contribution of the IR modes vanishes identically . As it turns out  contact terms yield only a
 toplogical Chern-Simons Lagrangian which in two dimensions vanishes identically . This also reflects in a sense the fact that the above UV modes are suppressed in the
 propagator $({\cal D}_a)^{-1}$. On the other hand although the first term $\frac{a{\Gamma}^{L}}{2}$ in (\ref{result0})
 vanishes as it stands in the continuum limit , its contribution in that limit is not zero and is exactly given by the
 canonical theta term. Putting these facts together we conclude that the anomaly emerges essentially from the UV region
 of the spectrum as expected. We now give a rigorous proof of this result.

In the presence of gauge fields the free expansion (\ref{result0}) becomes a covariant expansion given by
\begin{eqnarray}
\frac{1}{{\cal D}_{AF}}=\frac{1}{{\cal D}_{GF}^2}\frac{1}{a}\bigg(\frac{1}{2}\{{\Gamma}^R,\hat{\Gamma}^L\}\frac{1}{\hat{\Gamma}^{L2}}+\frac{1}{2}[{\Gamma}^R,\hat{\Gamma}^L]\frac{1}{\hat{\Gamma}^{L2}}+1\bigg)\hat{\Gamma}^L.
\end{eqnarray}
We use now the results
\begin{eqnarray}
&&\{{\Gamma}^R,\hat{\Gamma}^L\}=a^2\bigg[{\cal
D}_{AF}^2-F^F-\frac{2}{a^2}\bigg]~,~
[{\Gamma}^R,\hat{\Gamma}^L]=2iab{\cal D}_{GF}^{'}.
\end{eqnarray}
$F^F$ is defined in (\ref{ginsparg-wilson}) and ${\cal
D}_{AF}^{'}$ is the gauged Watamuras Dirac operator defined by $
{\cal D}_{AF}^{'}={\epsilon}_{abc}{\sigma}_ax_b^FL_c^R$ where
$x_b^F$'s are the covariant coordinates given by
$x_b^F=\frac{D_b^F}{\sqrt{l(l+1)}}$. In other words $x_b^F$
reduces to $n_b^F$ in the absence of gauge fields and to $n_b$ in
the continuum limit. We use also the fact that
${\hat{\Gamma}}^L=b{\sigma}_ax_a^F+\frac{a}{2}$ to deduce the
result $\hat{\Gamma}^{L2}=1+a^2F^F$ , then by putting all these
ingredients together we find that the gauge covariant expansion
of the full propagator is given by
\begin{eqnarray}
&&\frac{1}{{\cal
D}_{AF}}=\frac{a\hat{\Gamma}^L}{2}-\frac{a^3}{2}\frac{F^F}{1+a^2F^F}\hat{\Gamma}^L+\frac{1}{{\cal
D}_{Aa}},\label{result1}
\end{eqnarray}
with
\begin{eqnarray}
&&\frac{1}{{\cal D}_{Aa}}=\frac{1}{{\cal D}^2_{AF}}\bigg[ib{\cal
D}_{AF}^{'}+\frac{a}{2}F^F\bigg]\frac{1}{1+a^2F^F}\hat{\Gamma}^L.\label{result2}
\end{eqnarray}
It is not difficult to see that each term in this expansion is exactly covariant under gauge transformations . It is also obvious that (\ref{result1}) reduces in the limit $A^F_a{\longrightarrow}0$ to (\ref{result0}) . Furthermore in the continuum limit $l{\longrightarrow}{\infty}$ , (\ref{result1}) reduces to $\frac{1}{{\cal D}_{A}}=\frac{1}{{\cal D}_{A}^2}(i{\cal D}_{A}^{'}{\gamma})$ which is actually an identity since ${\cal D}_A=i{\cal D}_A^{'}{\gamma}$ and thus the expansion (\ref{result1}) is simply not available to us in the continuum .

In order to compute the  contribution of $({\cal D}_{Aa})^{-1}$ it
is obvious that one needs also an expansion of $\frac{1}{{\cal
D}_{AF}^2}$ . To this end we recall first that the square of the
Dirac operator ${\cal D}_{AF}$ is given by ${\cal D}_{AF}^2={\cal
D}_{AF}+({L}_a-L_a^R+{A}_a^F)^2+\frac{i}{2}{\epsilon}_{abc}{\sigma}_cF_{ab}^F$
. By using the constraint (\ref{local}) we can write this square
in the form
\begin{eqnarray}
&&{\cal
D}_{AF}^2=2l(l+1){P}_{AF}\bigg[1-\frac{1}{{P}_{AF}}v_a^Fn_a^R\bigg],\label{covexp}
\end{eqnarray}
where $n_a^R$'s are the fuzzy coordinates which act from the right
, i.e $n_a^R=\frac{L_a^R}{\sqrt{l(l+1)}}$  and where
\begin{eqnarray}
&&{P}_{AF}=1+\frac{1+F^F+\sqrt{l(l+1)}{\sigma}x^F}{2l(l+1)}~,~
v_a^F=x_a^F+\frac{{\sigma}_a}{2\sqrt{l(l+1)}}.\label{covexp10}
\end{eqnarray}
It is clear that the only bit in ${\cal D}_{AF}^2$ which acts on
the right is $\vec{n}^R$ and that $P_{AF}$ acts entirely on the
left . This means that the operator $(P_{AF})_{\alpha \beta}$ can
now be treated as a matrix in $Mat_{2l+1}$ . We think now of
${P}_{AF}$ as a propagator and of $n_a^Rv_a^F$ as a vertex and
write the expansion

\begin{eqnarray}
\frac{1}{{\cal D}_{AF}^2}&=&\frac{1}{2l(l+1)}\sum_{N=0}^{\infty}\bigg(\frac{1}{{P}_{AF}}v_a^Fn_a^R\bigg)^N\frac{1}{{P}_{AF}}\nonumber\\
&=&\frac{1}{2l(l+1)}\sum_{N=0}^{\infty}\bigg(\frac{1}{{P}_{AF}}v_{a_1}^F\frac{1}{{P}_{AF}}v_{a_2}^F...\frac{1}{{P}_{AF}}v_{a_N}^F\frac{1}{{P}_{AF}}\bigg)\bigg(n_{a_1}^Rn_{a_2}^R...n_{a_N}^R\bigg).\label{covexp00}\nonumber\\
\end{eqnarray}
The meaning of this expansion will only be clear in the continuum
limit which we will  take shortly. The propagator
$\frac{1}{{P}_{AF}}$ acts now from the left and thus it can also
be thought of as a matrix in $Mat_{2l+1}{\otimes}Mat_{2}$ rather
than as an operator . It is given by
\begin{eqnarray}
\frac{1}{{P}_{AF}}=(1-\frac{a}{2}\hat{\Gamma}^L-\frac{{a}^2}{2}F^F)\frac{1}{1-\frac{{a}^3}{4-{a}^2}\{\hat{\Gamma}^L,F^F\}
-\frac{{a}^4}{4-{a}^2}(F^{F})^2},\label{covexp}
\end{eqnarray}
Therefore given any two operators $(X_a^L)_{\alpha \beta}$ and
$Y_b^R$ which act from the left and from the right respectively ,
i.e $(X_a^L)_{\alpha \beta}f{\equiv}(X_a)_{\alpha \beta}f$ and
$Y_b^Rf{\equiv}fY_b$ for any $f{\in}Mat_{2l+1}$ ,  we can compute
\begin{eqnarray}
\bigg(\frac{1}{{\cal D}_{AF}^2}X_a^LY_b^R\bigg)_{{\alpha
\beta}}^{AB,CD}=\frac{1}{2l(l+1)}\sum_{N=0}^{\infty}\bigg(\frac{1}{{P}_{AF}}v_{a_1}^F\frac{1}{P_{AF}}v_{a_2}^F...
\frac{1}{{P}_{AF}}v_{a_N}^F\frac{1}{{P}_{AF}}X_a\bigg)_{\alpha
\beta}^{AC}\bigg(Y_bn_{a_N}^F...n_{a_2}^Fn_{a_1}^F\bigg)^{DB}.\label{covexp1}\nonumber\\
\end{eqnarray}
To summarize , the gauge covariant expansion of the propagator
$\frac{1}{{\cal D}_{AF}}$  is defined by  equations
(\ref{result1}) , (\ref{result2}), (\ref{covexp00}) and
(\ref{covexp}) . Using these equations together with the
definition (\ref{covexp1}) one obtains an explicit formula for
the exact  quark propagator on ${\bf S}^2_F$ , viz

\begin{eqnarray}
&&\bigg(\frac{1}{{\cal D}_{AF}}\bigg)^{AB,CD}_{\alpha \beta}
=\bigg(\frac{a\hat{\Gamma}^L}{2}-\frac{a^3}{2}\frac{F^F}{1+a^2F^F}\hat{\Gamma}^L\bigg)^{AC}_{\alpha
\beta}{\delta}^{BD} +\bigg(\frac{1}{{\cal
D}_{Aa}}\bigg)^{AB,CD}_{\alpha \beta},\label{result11}
\end{eqnarray}
with
\begin{eqnarray}
\bigg(\frac{1}{{\cal D}_{Aa}}\bigg)_{\alpha
\beta}^{AB,CD}&=&\frac{ib{\epsilon}_{abc}}{2l(l+1)}\sum_{N=0}^{\infty}\bigg(\frac{1}{{P}_{AF}}v_{a_1}^F\frac{1}{P_{AF}}v_{a_2}^F...
\frac{1}{{P}_{AF}}v_{a_N}^F\frac{1}{{P}_{AF}}{\sigma}_ax_b^F\frac{1}{1+a^2F^F}{\hat{\Gamma}}^L\bigg)_{\alpha
\beta}^{AC}\bigg(L_cn_{a_N}^F...n_{a_2}^Fn_{a_1}^F\bigg)^{DB}\nonumber\\
&+&\frac{a}{4l(l+1)}\sum_{N=0}^{\infty}\bigg(\frac{1}{{P}_{AF}}v_{a_1}^F\frac{1}{P_{AF}}v_{a_2}^F...
\frac{1}{{P}_{AF}}v_{a_N}^F\frac{1}{{P}_{AF}}\frac{F^F}{1+a^2F^F}{\hat{\Gamma}}^L\bigg)_{\alpha
\beta}^{AC}\bigg(n_{a_N}^F...n_{a_2}^Fn_{a_1}^F\bigg)^{DB}.\label{result22}\nonumber\\
\end{eqnarray}
As we have explained this gauge covariant expansion does not
exist on the continuum sphere . It will be used here to derive the
fuzzy axial anomaly and its continuum limit
 the
local axial anomaly .

\subsection{Local Axial Anomaly }
This expansion isolates in fact ( in a gauge covariant fashion )
the anomalous bit in the fermion propagator . More precisely ,
although the first term in (\ref{result11}) vanishes as it stands
in the large $l$ limit , its contribution ( i.e its trace ) gives
exactly the canonical anomaly . Indeed we can easily compute

\begin{eqnarray}
{\delta}_1{\cal A}_{{\theta}F} &\equiv
&-\frac{i}{2l+1}{\epsilon}_{abc}({\lambda}^FF_{ab}^F)^{BD}\bigg[tr_2{\sigma}_c\bigg(\frac{a\hat{\Gamma}^L}{2}\bigg)^{DA,BA}+tr_2{\sigma}_c
\bigg(-\frac{a^3}{2}\frac{F^F}{1+{a}^2F^F}\hat{\Gamma}^L\bigg)^{DA,BA}\bigg]\nonumber\\
&=&-2b\frac{i{\epsilon}_{abc}}{2l+1}Tr_l{\lambda}^FF_{ab}^F{\delta}_1x_c^F,\label{result3}
\end{eqnarray}
where
\begin{eqnarray}
{\delta}_1x_c^F&=&x_c^F-\frac{{a}^2}{2b}tr_2\bigg({\sigma}_c\frac{F^F}{1+{a}^2F^F}\hat{\Gamma}^L\bigg).\label{result33}
\end{eqnarray}
This gauge-covariant vector $\vec{\delta}_1x^F$ in the continuum limit becomes exactly the unit vector $\vec{n}$ on the $2-$dimensional sphere and hence the fuzzy anomaly (\ref{result3}) reduces in that limit to the theta term on ${\bf S}^2$ , in other words
\begin{eqnarray}
-2b\frac{i{\epsilon}_{abc}}{2l+1}Tr_l{\lambda}^FF_{ab}^F{\delta}_1x_c^F{\longrightarrow}-2i{\epsilon}_{abc}\int_{{\bf
S}^2}
\frac{d{\Omega}}{4{\pi}}{\lambda}(\vec{n})F_{ab}(\vec{n})n_c~,~{\rm
when}~l{\longrightarrow}{\infty}.
\end{eqnarray}
For example by using the star product on ${\bf S}^2_F$ one can
see that in the continuum limit $l{\longrightarrow}{\infty}$
where $a{\longrightarrow}0$ and $b{\longrightarrow}1$ we have
${\delta}_1x^F_a{\longrightarrow}n_a$ , $F_{ab}^F{\longrightarrow}F_{ab}$ ,
${\lambda}^F{\longrightarrow}{\lambda}$ while the trace
$\frac{1}{2l+1}Tr_l$ becomes the integral $\int_{{\bf S}^2}
\frac{d{\Omega}}{4{\pi}}$ .

\subsection{The Chern-Simons Action}

From equation (\ref{result2}) one can immediately read the
remaining two extra corrections corresponding to  the propagator
$\frac{1}{{\cal D}_{Aa}}$ , viz
\begin{eqnarray}
{\delta}_2{\cal A}_{\theta
F}&=&-\frac{i}{2l+1}{\epsilon}_{abc}({\lambda}^F
F_{ab}^F)^{BD}tr_2{\sigma}_c
\bigg(\frac{1}{{\cal D}_{AF}^2}(ib{\cal D}_{AF}^{'})\frac{1}{1+{a}^2F^F}\hat{\Gamma}^L\bigg)^{DA,BA}\nonumber\\
&=&-i{\epsilon}_{abc}\bigg(\frac{1}{1+{a}^2F^F}\hat{\Gamma}^L{\lambda}^FF_{ab}^F{\sigma}_c\bigg)_{\beta
\alpha}^{CD}\bigg(\frac{1}{{\cal
D}_{AF}^2}(\frac{i}{2}{b}^2{\epsilon}_{pqr}{\sigma}_px_q^Fn_r^R)\bigg)^{DA,CA}_{\alpha
\beta},
\end{eqnarray}
and
\begin{eqnarray}
{\delta}_3{\cal A}_{\theta F}&=&-\frac{i}{2l+1}{\epsilon}_{abc}({\lambda}^F F_{ab}^F)^{BD}tr_2{\sigma}_c\bigg(\frac{1}{{\cal D}_{AF}^2}(\frac{a}{2} F^F)\frac{1}{1+{a}^2F^F}\hat{\Gamma}^L\bigg)^{DA,BA}\nonumber\\
&=&-i{\epsilon}_{abc}\bigg(\frac{1}{1+a^2F^F}\hat{\Gamma}^L{\lambda}^FF_{ab}^F{\sigma}_c\bigg)_{\beta
\alpha}^{CD}\bigg(\frac{1}{{\cal
D}_{AF}^2}(\frac{{a}^2}{4}F^F)\bigg)^{DA,CA}_{\alpha \beta}.
\end{eqnarray}
By using the expansion (\ref{covexp00}) with (\ref{covexp})
together with the definition (\ref{covexp1}) we obtain for
${\delta}_2{\cal A}_{\theta F}$ and ${\delta}_3{\cal A}_{\theta
F}$ the formulae
\begin{eqnarray}
{\delta}_2{\cal A}_{\theta F}&=&-2b\frac{i{\epsilon}_{abc}}{2l+1}Tr_l{\lambda}^FF_{ab}^F{\delta}_2x_{c}^{F}\nonumber\\
{\delta}_2x_{c}^{F}&=&\sum_{N=1}\bigg(\frac{1}{2l+1}Tr_l(n_{a_N}^F...n_{a_1}^Fn_r^F)\bigg)\bigg(tr_2\big(\frac{1}{{P}_{AF}}v_{a_1}^F\frac{1}{{P}_{AF}}v_{a_2}^F...\frac{1}{{P}_{AF}}v_{a_N}^F{\delta}^F_{rc}\big)\bigg)\label{result5}
\end{eqnarray}
and
\begin{eqnarray}
{\delta}_3{\cal A}_{\theta F}&=&-2{b}\frac{i{\epsilon}_{abc}}{2l+1}Tr_l{\lambda}^FF_{ab}^F{\delta}_3x_{c}^{F}\nonumber\\
{\delta}_3x_{c}^{F}&=&\sum_{N=0}\bigg(\frac{1}{2l+1}Tr_l(n_{a_N}^F...n_{a_1}^F)\bigg)\bigg(tr_2\big(\frac{1}{{P}_{AF}}v_{a_1}^F\frac{1}{{P}_{AF}}v_{a_2}^F...\frac{1}{{P}_{AF}}v_{a_N}^F{\delta}_c^F\big)\bigg),\label{result6}
\end{eqnarray}
where ${\delta}^F_{rc}$ and ${\delta}_c^F$ stand for
\begin{eqnarray}
{\delta}_{rc}^F=\frac{i{\epsilon}_{rpq}}{2b}\frac{1}{{P}_{AF}}{\sigma}_px_q^F\frac{1}{1+{a}^2F^F}\hat{\Gamma}^L{\sigma}_c~,~{\delta}_c^F=\frac{{a}^2}{4{b}^3}\frac{1}{{P}_{AF}}F^F\frac{1}{1+{a}^2F^F}\hat{\Gamma}^L{\sigma}_c.
\end{eqnarray}
Before we take the continuum limit of these expressions we remark
that the components of the two vectors $\vec{\delta}_2x^{F}$ and
$\vec{\delta}_3x^{F}$ are matrices in $Mat_{2l+1}$ which are
covariant under gauge transformations. Furthermore because of the
presence of the traces $Tr_l(n_{a_N}^F...n_{a_1}^Fn_r^F)$ and
$Tr_l(n_{a_N}^F...n_{a_1}^F)$ , the components
${\delta}_2x_c^{F}$ and ${\delta}_3x_{c}^{F}$ are also  invariant
under the extra symmetry
$n^F_a{\longrightarrow}w_a^F=W^{+}n_a^FW$ where $W$ is an
arbitrary unitary transformation given by $W=e^{i{\alpha}^F}$,
${\alpha}^F{\in}Mat_{2l+1}$. These transformations $W$ are different
from gauge transformations $U$ introduced in section $3$ . For
example one can see from their explicit action
\begin{eqnarray}
n_a^F{\longrightarrow}w_a^F=W^{+}n_a^FW=n_a^F+\frac{1}{\sqrt{l(l+1)}}W^{+}[L_a,W],\label{514}
\end{eqnarray}
that the fuzzy coordinates $w_a^F$ reduce to the same continuum
coordinates $n_a$ in the limit, in other words this extra symmetry
is not available in the continuum . Using this symmetry one can
therefore rewrite the vectors ${\delta}_2x_c^{F}$ and
${\delta}_3x_c^{F}$ in the equivalent form
\begin{eqnarray}
{\delta}_2x_{c}^{F}&=&\sum_{N=1}\bigg(\frac{1}{2l+1}Tr_l(w_{a_N}^F...w_{a_1}^Fw_r^F)\bigg)\bigg(tr_2\big(\frac{1}{{P}_{AF}}v_{a_1}^F\frac{1}{{P}_{AF}}v_{a_2}^F...\frac{1}{{P}_{AF}}v_{a_N}^F{\delta}_{rc}^F\big)\bigg)\label{result5prime}
\end{eqnarray}
and
\begin{eqnarray}
{\delta}_3x_{c}^{F}&=&\sum_{N=0}\bigg(\frac{1}{2l+1}Tr_l(w_{a_N}^F...w_{a_1}^F)\bigg)\bigg(tr_2\big(\frac{1}{{P}_{AF}}v_{a_1}^F\frac{1}{{P}_{AF}}v_{a_2}^F...\frac{1}{{P}_{AF}}v_{a_N}^F{\delta}_c^F\big)\bigg).\label{result6prime}
\end{eqnarray}
We can now immediately write down the continuum limit of the
above expressions . We already know that in this limit
$a{\longrightarrow}0$ , $b{\longrightarrow}1$ ,
$\frac{1}{2l+1}Tr_l{\longrightarrow}\int_{{\bf S}^2}$ ,
$n^F_a{\longrightarrow}n_a$ , $x_a^F{\longrightarrow}n_a$,
$F_{ab}^F{\longrightarrow}F_{ab}$ ,
${\lambda}^F{\longrightarrow}{\lambda}$ , ${\alpha}^F{\longrightarrow}{\alpha}$ and
$\hat{\Gamma}^L{\longrightarrow}{\gamma}$ . From equation
(\ref{covexp10}) we also see that in this limit
${{P}_{AF}}{\longrightarrow}1$ and $v_a^F{\longrightarrow}n_a$ ,
whereas from equation (\ref{514}) we see that
$w_a^F{\longrightarrow}n_a$ . For example the continuum limit of
(\ref{result5}) is given by
\begin{eqnarray}
{\delta}_2{\cal A}_{\theta}&=&-2i{\epsilon}_{abc}\int_{{\bf S}^2}\frac{d{\Omega}}{4{\pi}}{\lambda}(\vec{n})F_{ab}(\vec{n}){\delta}_2x_c^{}(\vec{n})\nonumber\\
{\delta}_2x_c^{}(\vec{n})&=&\frac{i}{2}\int_{{\bf
S}^2}\frac{d{\Omega}^{'}}{4{\pi}}(\vec{n}{\times}\vec{n}^{'})_ptr_2\bigg({\sigma}_c
\sum_{N=1}(\vec{n}.\vec{n}^{'})^N{\sigma}_p{\gamma}(\vec{n})\bigg).
\end{eqnarray}
It is easily seen that all the expected continuum divergence
arises from the series $\sum_{N=1}(\vec{n}.\vec{n}^{'})^N$ when
$\vec{n}=\vec{n}^{'}$ , i.e from contact terms . Thus we
need first to separate contact terms as follows
\begin{eqnarray}
{\delta}_2x_c^{}(\vec{n})&=&\frac{i}{2}\bigg[(\vec{n}{\times}\vec{n}^{'})_ptr_2\bigg({\sigma}_c
\sum_{N=1}(\vec{n}.\vec{n}^{'})^N{\sigma}_p{\gamma}(\vec{n})\bigg)\bigg]_{\vec{n}^{'}=\vec{n}}+{\delta}_2x_c(\vec{n})|_{nct}
\label{result7}
\end{eqnarray}
where
\begin{eqnarray}
{\delta}_2x_c(\vec{n})|_{nct}&=&\frac{i}{2}\int_{\vec{n}^{'}{\neq}\vec{n}}\frac{d{\Omega}^{'}}{4{\pi}}(\vec{n}{\times}\vec{n}^{'})_ptr_2\bigg({\sigma}_c
\sum_{N=1}(\vec{n}.\vec{n}^{'})^N{\sigma}_p{\gamma}(\vec{n})\bigg).
\end{eqnarray}
It is clear that in ${\delta}_2x_c^F(\vec{n})|_{nct}$ we have $|\vec{n}.\vec{n}^{'}|=|cos{\theta}|<1$ where $\theta$ is the
angle between $\vec{n}$ and $\vec{n}^{'}$ and thus the series
$\sum_{N=1}(\vec{n}.\vec{n}^{'})^N$ is now convergent . We can therefore
simply substitute the result
\begin{eqnarray}
\sum_{N=1}(\vec{n}.\vec{n}^{'})^N=\frac{\vec{n}.\vec{n}^{'}}{1-\vec{n}.\vec{n}^{'}},\nonumber
\end{eqnarray}
and thus obtain
\begin{eqnarray}
{\delta}_2x_c(\vec{n})|_{nct}&=&\int_{\vec{n}^{'}{\neq}\vec{n}}\frac{d{\Omega}^{'}}{4{\pi}}\bigg(n_c(\vec{n}.\vec{n}^{'})-n_c^{'}\bigg)\frac{\vec{n}.\vec{n}^{'}}{1-\vec{n}.\vec{n}^{'}}\nonumber\\
&=&(n_cn_d-{\delta}_{cd})n_e\int_{{\bf S}^2}\frac{d{\Omega}^{'}}{4{\pi}}\frac{n_d^{'}{n}^{'}_e}{1+{\epsilon}-\vec{n}.\vec{n}^{'}},\nonumber
\end{eqnarray}
where in the second line we have replaced
$\int_{\vec{n}^{'}{\neq}\vec{n}}\frac{d{\Omega}^{'}}{4{\pi}}$ by
the intgeral over the full sphere $\int_{{\bf
S}^2}\frac{d{\Omega}^{'}}{4{\pi}}$ with the prescription
${\epsilon}>0$ so that the contribution of the terms
$\vec{n}^{'}=\vec{n}$ is kept equal to zero . 

Remark that $n_c{\delta}_2x_c(\vec{n})|_{nct}=0$ and hence the contribution of the above non-contact terms is identically zero since in the continuum we also have $n_aA_a=0$ , ${\epsilon}_{abc}F_{ab}=2in_c{\partial}_b(A_b)$ and thus
\begin{eqnarray}
{\delta}_2{\cal A}_{\theta}|_{nct}&=&-2i{\epsilon}_{abc}\int_{{\bf S}^2}\frac{d{\Omega}}{4{\pi}}{\lambda}(\vec{n})F_{ab}(\vec{n}){\delta}_2x_c^{}(\vec{n})|_{nct}\nonumber\\
&=&4\int_{{\bf S}^2}\frac{d{\Omega}}{4{\pi}}{\lambda}{\partial}_b(A_b)n_c{\delta}_2x_c|_{nct}\nonumber\\
&{\equiv}&0.	
\end{eqnarray}
For contact terms the series $\sum_{N=1}(\vec{n}.\vec{n}^{'})^N$
is obviously divergent and thus the first term of equation
(\ref{result7})  is not well defined . Thus it is natural to go back to the finite
expression (\ref{result5prime}) and think of it now as a gauge-invariant
regularization of  contact terms with a cut-off $l$ provided here by the
fuzzy sphere . Since we are interested
in the continuum theory the cut-off $l$ is effectively large and thus one may retain only the first few corrections to the continuum theory . As it turns
out the contribution of contact terms is regularized completely and in a gauge-invariant manner if we only keep $O(\frac{1}{l})$
corrections. To see how all this works
explicitly , we use the star product on ${\bf S}^2_F$ and replace
${\delta}_2x_c^F$ by its image $<{\delta}_2x_c^F>(\vec{n})$ via
the coherent states , namely
\begin{eqnarray}
<{\delta}_2x_{c}^{F}>(\vec{n})&=&\sum_{N=1}\bigg(\int_{{\bf
S}^2}\frac{d{\Omega}^{'}}{4{\pi}}
<w_{a_N}^F>*...*<w_{a_1}^F>*<w_r^F>(\vec{n}^{'})\bigg)\nonumber\\
&{\times}&\bigg(tr_2<\frac{1}{{P}_{AF}}>*<v_{a_1}^F>*<\frac{1}{{P}_{AF}}>*<v_{a_2}^F>*...*<\frac{1}{{P}_{AF}}>*<v_{a_N}^F>*<{\delta}_{rc}^F>(\vec{n})\bigg).\nonumber
\end{eqnarray}
Since non-contact terms are regular in the limit one can still
treat them in the same way as before and hence their contribution
is zero . This means that $<{\delta}_2x_{c}^{F}>(\vec{n})$ is
dominated by contact terms , namely

\begin{eqnarray}
<{\delta}_2x_{c}^{F}>(\vec{n})&=&\sum_{N=1}\bigg(
<w_{a_N}^F>*...*<w_{a_1}^F>*<w_r^F>(\vec{n})\bigg)\nonumber\\
&{\times}&\bigg(tr_2<\frac{1}{{P}_{AF}}>*<v_{a_1}^F>*<\frac{1}{{P}_{AF}}>*<v_{a_2}^F>*...*<\frac{1}{{P}_{AF}}>*<v_{a_N}^F>*<{\delta}_{rc}^F>(\vec{n})\bigg).\nonumber
\end{eqnarray}
From equation (\ref{2.13}) we have 
$<n^F_a>=n_a-\frac{1}{2l}n_a+O(\frac{1}{l^2})$ and thus we can compute to
this order in $\frac{1}{l}$ the following quantities
\begin{eqnarray}
&&<\hat{\Gamma}^L>={\gamma}+\frac{1}{l}(-\frac{1}{2}{\gamma}+{\sigma}A+\frac{1}{2})+O(\frac{1}{l^2})~,~<\frac{1}{{P}_{AF}}>=1-\frac{1}{2l}{\gamma}+O(\frac{1}{l^2})\nonumber\\
&&<v_a^F>=n_a+\frac{1}{l}(-\frac{1}{2}n_a+A_a+\frac{{\sigma}_a}{2})+O(\frac{1}{l^2})~,~<w_a^F>=n_a+\frac{1}{l}(-\frac{1}{2}n_a+i{\cal
L}_a(\alpha))+O(\frac{1}{l^2})\nonumber\\
&&<{\delta}_{rc}^F>=\frac{i}{2}{\epsilon}_{rpq}\bigg(n_q{\sigma}_p{\gamma}{\sigma}_c+\frac{1}{l}{\sigma}_p(-n_q{\gamma}+n_q{\sigma}A+\frac{1}{2}n_q+{\gamma}A_q){\sigma}_c-\frac{1}{2l}n_q{\gamma}{\sigma}_p{\gamma}{\sigma}_c\bigg)+O(\frac{1}{l^2}).\nonumber
\end{eqnarray}
Also by using the star product on ${\bf S}^2_F$ we compute
\begin{eqnarray}
<\frac{1}{{P}_{AF}}>*<v_a^F>=<\frac{1}{{P}_{AF}}><v_a^F>+O(\frac{1}{l^2})=n_a+\frac{1}{l}\bigg(-\frac{1}{2}n_a+A_a+\frac{{\sigma}_a}{2}-\frac{1}{2}{\gamma}n_a\bigg)+O(\frac{1}{l^2}).\nonumber
\end{eqnarray}
Remark for example the continuum limit
${\delta}_{rc}^F{\longrightarrow}{\delta}_{rc}=\frac{i}{2}{\epsilon}_{rpq}{\sigma}_pn_q{\gamma}{\sigma}_c$ when $l{\longrightarrow}{\infty}$ as well as
 the limit $<\frac{1}{{P}_{AF}}>*<v_a^F>{\longrightarrow}n_a$ when $l{\longrightarrow}{\infty}$. Finally and by
using once again the star product we compute the formulae
\begin{eqnarray}
<w_{a_N}^F>*...*<w_{a_1}^F>*<w_r^F>(\vec{n})&=&<w_{a_N}^F>...<w_{a_1}^F><w_r^F>\nonumber\\
&+&\frac{1}{2l}K_{pq}n_r\sum_{i=2}^N\sum_{j<i}n_{a_N}...{\partial}_{p}n_{a_i}...{\partial}_{q}n_{a_j}...n_{a_1}\nonumber\\
&+&\frac{1}{2l}K_{pr}{\partial}_{p}(n_{a_N}...n_{a_1})+O(\frac{1}{l^2}),
\nonumber
\end{eqnarray}
and
\begin{eqnarray}
<\frac{1}{{P}_{AF}}>*<v_{a_1}^F>*...*<\frac{1}{{P}_{AF}}>*<v_{a_N}^F>*<{\delta}_{rc}^F>&=& <\frac{1}{{P}_{AF}}><v_{a_1}^F>...<\frac{1}{{P}_{AF}}><v_{a_N}^F><{\delta}_{rc}^F>\nonumber\\
&+&
\frac{1}{2l}{\delta}_{rc}K_{pq}\sum_{i=1}^{N-1}\sum_{j>i}n_{a_1}...{\partial}_{p}n_{a_i}...{\partial}_{q}n_{a_j}...n_{a_N}\nonumber\\
&+&
\frac{1}{2l}K_{pq}{\partial}_{p}(n_{a_1}...n_{a_N}){\partial}_{q}{\delta}_{rc}+O(\frac{1}{l^2}).
\nonumber
\end{eqnarray}
Putting all these results together we obtain the contribution of contact terms 
in the form
\begin{eqnarray}
<{\delta}_2x_{c}^{F}>(\vec{n})&=&tr_2\bigg(\sum_{N=1}\bigg(<\frac{1}{{P}_{AF}}><v_a^F><w_{a}^F>\bigg)^N<{\delta}_{rc}^F><w_r^F>\bigg)\nonumber\\
&=&(l+1)\bigg(tr_2<{\delta}_{rc}^F>\bigg)<w_r^F>,\nonumber
\end{eqnarray}
where we have also  used the result
\begin{eqnarray}
\sum_{N=1}\bigg(<\frac{1}{{P}_{AF}}><v_a^F><w_{a}^F>\bigg)^N=\sum_{N=1}(1-\frac{1}{l})^N=l+1.\nonumber
\end{eqnarray}
In other words the sum over $N$ is linearly divergent with $l$ and hence the remaining term $\big( tr_2<{\delta}_{rc}^F>\big)<w_r^F>$ must approach in the continuum limit $0$ at least as $\frac{1}{l}$ in order to reproduce a finite expression. A final calculation shows that this is indeed the case where from the equation
\begin{eqnarray}
tr_2<{\delta}_{rc}^F>&=&\frac{i}{2}{\epsilon}_{rpq}n_qtr_2{\sigma}_p{\gamma}{\sigma}_c+\frac{i{\epsilon}_{rpq}}{2l}tr_2{\sigma}_c{\sigma}_p\big(-n_q\gamma+n_q\sigma
A+\frac{1}{2}n_q+\gamma
A_q\big)-\frac{i{\epsilon}_{rpq}}{4l}n_qtr_2\gamma{\sigma}_p\gamma
{\sigma}_c\nonumber\\
&=&(n_rn_c-{\delta}_{rc})+\frac{i{\epsilon}_{rcq}n_q}{l}-\frac{1}{l}[n_rn_c-{\delta}_{rc}-n_cA_r-n_rA_c],\nonumber
\end{eqnarray}
we deduce that  $\big(tr_2<{\delta}_{rc}^F>\big)<w_r^F>=\frac{1}{l}[A_c-i{\cal L}_c(\alpha)]$ . The contribution of contact terms is therefore finite equal to
\begin{eqnarray}
<{\delta}_2x_{c}^{F}>(\vec{n})&=&A_c-i{\cal L}_c(\alpha),\label{alpha}
\end{eqnarray}
and as a consequence
\begin{eqnarray}
{\delta}_2{\cal A}_{\theta}=-2i{\epsilon}_{abc}\int_{{\bf S}^2}\frac{d{\Omega}}{4{\pi}}{\lambda}(\vec{n})F_{ab}(\vec{n})A_c(\vec{n}).\label{CS}
\end{eqnarray}
The dependence on the arbitrary function $\alpha$ drops because of the identity ${\epsilon}_{abc}F_{ab}{\cal L}_c(\alpha)=0$. This is equivalent to the fact that the term $-i{\cal L}_c(\alpha)$ in (\ref{alpha}) can always be canceled by a continuum gauge transformation $A_{a}^{'}=A_a-i{\cal L}_a(\Omega)$, i.e the fuzzy symmetries (\ref{514}) are not available as expected in the continuum setting since they can be eliminated by gauge symmetries . (\ref{CS}) is exactly the topological Chern-Simons action in two dimensions and thus it vanishes identically as one might easily check. Its emergence from contact terms of the propagator $\frac{1}{{\cal D}_{aA}}$ is intimately related to the fact that the top modes of the Dirac operator ${\cal D}_{Aa}$ as opposed to those of the exact Diarc operator ${\cal D}_{AF}$ are very large .

Similarly to above the last correction to the anomaly given by equation (\ref{result6}) has the limit
\begin{eqnarray}
{\delta}_3{\cal A}_{\theta}&=&-2i{\epsilon}_{abc}\int_{{\bf S}^2}\frac{d{\Omega}}{4{\pi}}{\lambda}(\vec{n})F_{ab}(\vec{n}){\delta}_3x_c^{}(\vec{n})\nonumber\\
{\delta}_3x_c^{}(\vec{n})&=&\int_{{\bf
S}^2}\frac{d{\Omega}^{'}}{4{\pi}}tr_2\bigg(
\sum_{N=0}(\vec{n}.\vec{n}^{'})^N{\delta}_c(\vec{n})\bigg).
\end{eqnarray}
The vector $\vec{\delta}(\vec{n})$ in above is in fact identically zero as one might easily check . Again for non-contact terms the series $\sum_{N=0}(\vec{n}.\vec{n}^{'})^N$ converges and thus the corresponding contribution clearly vanishes . For contact terms we have to regularize as before but since now the vector $<\vec{\delta}^F>(\vec{n})$ vanishes as $\frac{1}{l^2}$ and not as $\frac{1}{l}$ we immediately conclude that the contribution of contact terms is zero in this case and thus we obatin the result
\begin{eqnarray}
{\delta}_3{\cal A}_{\theta}&\equiv&0.
\end{eqnarray}

\section{Conclusion}
We  showed that the local axial anomaly in $2-$dimensions emerges
naturally if one postulates an underlying noncommutative fuzzy structure
of spacetime . As we have discussed in this article this result consists in three main parts : 

$1)$ We showed that the Dirac-Ginsparg-Wilson relation on ${\bf S}^2_F$ contains already at the classical level the anomaly in the form of an edge effect which under quantization becomes precisely the ``fuzzy'' $U(1)_A$ axial anomaly on the fuzzy sphere. 

$2)$ We derived  a novel gauge-covariant expansion of the quark propagator in the form $\frac{1}{{\cal D}_{AF}}=\frac{a\hat{\Gamma}^L}{2}+\frac{1}{{\cal D}_{Aa}}$ where
$a=\frac{2}{2l+1}$ is the lattice spacing on ${\bf S}^2_F$ , $\hat{\Gamma}^L$ is the covariant noncommutative chirality and
${\cal D}_{Aa}$ is an effective Dirac operator which has essentially the same IR spectrum
as ${\cal D}_{AF}$ but differes from it on the UV modes since the eigenvalues of ${\cal D}_{Aa}$ on the top modes are very large compared to those of ${\cal D}_{AF}$ . Most remarkably however is the fact that both operators share the same continuum limit and thus the above covariant expansion is not available in the continuum theory . 

$3)$ The first bit in this expansion $\frac{a\hat{\Gamma}^L}{2}$ although it vanishes as it stands in the continuum limit , its contribution to the anomaly is exactly the canonical
theta term. The contribution of the propagator $\frac{1}{{\cal D}_{Aa}}$ is on the other hand equal to the toplogical Chern-Simons action which in two dimensions is identically zero . In particular we have explicilty shown that beside the cut-off $l$ provided by  the star product of the fuzzy sphere itself there is no need to any extra regulator even while approaching the limit . 

Finally we have to note that a complete extension of the above results to the case of monopoles  is more or less straightforward after we identify correctly the corresponding bundle structure on ${\bf S}^2_F$ . As it turns out this is indeed possible and thus the extension can be made without much difficulty . The relevant detail will be however reported elsewhere. The computation of the effective action of the Schwinger model on  the fuzzy two-sphere, which will allow us on the other hand to probe the solvability of the model , 
will also be reported elsewhere .

\begin{center}
{\bf\large Acknowledgments}
\end{center}
I would like to thank  A.P.Balachandran , G.Immirzi , Denjoe O'Connor and P.Presnajder for their
critical comments while the work was in progress. This work was
supported in part by the DOE under contract number
DE-FG02-85ER40231.

\section{Appendix }

\paragraph{$1$}On continuum ${\bf S}^2$ exact chiral invariance of the free classical
action is expressed by the anticommutation relation
$\{{\gamma},{\cal D}\}=0$ which is in fact the limit
of the Ginsparg-Wilson relation (\ref{GWrelation3}). In the presence of gauge fields , the Ginsparg-Wilson relation (\ref{GWrelation3}) becomes (\ref{dicta}) and the edge effect seen there is exactly the source of the anomaly . We will now show 
that if we try instead to formulate chiral invariance on the fuzzy sphere without  edge effect then the action becomes complex and not gauge invariant and thus not quantizable in any canonical fashion .

As we have already said the continuum limit of the classical actions (\ref{fuzziii}) and (\ref{gaugeFaction}) are given respectively by (\ref{classiii}) and (\ref{contaction1}) . (\ref{contaction1}) was already shown to be gauge invariant, but
under the canonical continuum chiral transformations
${\psi}{\longrightarrow}{\psi}^{'}={\psi}+{\lambda}{\gamma}{\psi}$
, $
\bar{\psi}{\longrightarrow}\bar{\psi}^{'}=\bar{\psi}+{\lambda}\bar{\psi}{\gamma}$
, one can show that it is invariant only if the gauge field is
constrained to satisfy $n_aA_a=0$ because of the
identity ${\cal D}_A{\gamma}+{\gamma}{\cal D}_A=2\vec{A}.\vec{n}
$. From the continuum limit of (\ref{local}) it is
obvious that this constraint is satisfied and hence chiral
symmetry is maintained. However one wants also to formulate
chiral symmetry without the need to use any constraint on the
gauge field , indeed the action

\begin{eqnarray}
\int \frac{d{\Omega}}{4{\pi}}\bigg[\bar{{\psi}}{\cal D}
{\psi}+\bar{{\psi}}\hat{\sigma}_a{A}_a{{\psi}}\bigg].\label{2}
\end{eqnarray}
is strictly chiral invariant for arbitrary gauge configurations .
$\hat{\sigma}_a$ is the Clifford algebra projected onto the sphere
, i.e $\hat{\sigma}_a={\cal P}_{ab}{\sigma}_{b}$ , ${\cal
P}_{ab}={\delta}_{ab}-n_an_b$ . The action (\ref{2}) is also
still gauge invariant because of the identity $n_a{\cal L}_a=0$ .
Action (\ref{2}) can be rewritten as follows
\begin{eqnarray}
S_{C}&=&\int \frac{d{\Omega}}{4{\pi}}\bigg[\bar{{\psi}}{\cal D}
{\psi}+{\epsilon}_{abc}\bar{{\psi}}Z_bn_c{A}_a{{\psi}}\bigg]~,~Z_a=\frac{i}{2}[{\gamma},{\sigma}_a].\label{3}
\end{eqnarray}
From this form of the action one can immediately define the following  chiral-covariant Dirac operator 
${\cal D}_C={\cal D}+{\epsilon}_{abc}Z_bn_cA_a$. The difference
between the continuum gauge-covariant Dirac operator ${\cal D}_A$
and the continuum chiral-covariant Dirac operator ${\cal D}_C$ is
proportional to the normal component ${\phi}=n_aA_a$ of the gauge field
, i.e ${\cal D}_{C}={\cal D}_A-{\gamma}{\phi}$ , and hence they
are essentially identical ( by virtue of the constraint $n_aA_a=0$
) , i.e (\ref{contaction1}) and (\ref{3}) are equivalent actions . The Fuzzy analogue of (\ref{contaction1}) is (\ref{gaugeFaction}) whereas the fuzzy analogue of (\ref{3}) is
the action
\begin{eqnarray}
S_{CF} =\frac{1}{2l+1}Tr_{l}
\bigg[\bar{{\psi}_F}{\cal D}_F{\psi}_F+{\epsilon}_{abc}\bar{{\psi}_F}Z_b^Fn^F_c{A}_{a}^F{{\psi}_F}
\bigg]~,~Z_a^F=\frac{i}{2}[{\Gamma}^L{\sigma}_a+{\sigma}_a{\Gamma}^R].\label{3F}
\end{eqnarray}
The underlying fuzzy Dirac operator  is obviously given by $
{\cal D}_{CF}={\cal D}_F+{\epsilon}_{abc}Z_b^Fn^F_c{A}_{a}^F~$ which
tends in the large
$l$ limit to ${\cal D}_C$ . This means in particular that the two Dirac operators ${\cal D}_{CF}$ and
${\cal D}_{AF}$ have the same continuum limit. As a consequence the corresponding fuzzy classical actions (\ref{3F}) and (\ref{gaugeFaction}) approach the same continuum action (\ref{contaction1}) in the limit.  Remark also that the Dirac operator ${\cal D}_{CF}$ satisfies the Ginsparg-Wilson relation ${\cal D}_{CF}{\Gamma}^R-{\Gamma}^L{\cal D}_{CF}=0$ and hence  the action (\ref{3F}) is invariant under free fuzzy chiral transformations
\begin{eqnarray}
&&{\psi}_F{\longrightarrow}{\psi}_F^{'}={\psi}_F+{\Gamma}^R{\psi}_F{\lambda}^F+O(({\lambda}^F)^2)\nonumber\\
&&\bar{\psi}_F{\longrightarrow}\bar{\psi}_F^{'}=\bar{\psi}_F-{\lambda}^F\bar{\psi}_F{\Gamma}^L+O(({\lambda}^F)^2).\label{Fchiral}
\end{eqnarray}
In here the chiral parameter ${\lambda}^F$ which is still 
 a general $(2l+1){\times}(2l+1)$ matrix  does not transform under gauge symmetries in contrast with  the case considered in section $4$. Also these fuzzy chiral transformations (\ref{Fchiral}) as opposed to the fuzzy chiral transformations (\ref{fuzzytrans2}) do not depend on the gauge field , yet both (\ref{Fchiral}) and (\ref{fuzzytrans2}) reduce in the limit to the same continuum chiral transformations . 

Remark however that the Dirac operator ${\cal D}_{CF}$ is not self-adjoint as well as not gauge-covariant. In other words the action (\ref{3F}) is complex and not gauge invariant and thus it is not suited for any quantization procedure.

\paragraph{$2$ Measure-Transforming Chiral Symmetries}
In the remainder of this appendix we introduce measure-changing chiral transformations as opposed to the  chiral transformations (\ref{fuzzytrans2}) which leave the measure invariant. These new transformations as it turns out yield the same anomaly (\ref{ep}) and thus they are equivalent to the transformations (\ref{fuzzytrans2}) . The only difference is the fact that this anomaly (\ref{ep}) arises now from the measure instead. We will also define gauge-invariant axial current as opposed to the gauge-covriant axial current defined in section $4.3$.

Remark that since ${\Gamma}^R-\hat{\Gamma}^L{\longrightarrow}-2{\gamma}$ when $l{\longrightarrow}{\infty}$ , the  continuum limit of (\ref{anomaly}) is the usual formal answer, i.e
\begin{eqnarray}
S_{{\theta}}&=&-\int \frac{d{\Omega}}{4{\pi}}{\lambda}(x)\sum_{\mu}{\phi}_{\mu}^{+}(x)(-2{\gamma}){\phi}_{\mu}(x).\label{anomalylimit}
\end{eqnarray}
In the continuum the sum $\sum_{\mu}$ is not cutoff since all orbital angular momenta are allowed and obviously ${\phi}_{\mu}(x)$ stands now for the eigenfunctions of the continuum gauged Dirac operator ${\cal D}_A={\cal D}+{\sigma}_aA_a$ . These states satisfy the completeness relation
\begin{eqnarray}
\sum_{\mu}{\phi}_{\alpha}(x){\phi}^{+}_{\beta}(y)={\delta}^2(x-y){\delta}_{\alpha \beta}.
\end{eqnarray}
If we try now to use this completeness relation in (\ref{anomalylimit}) we will instead get the ill-defined expression $S_{\theta}=2\int \frac{d{\Omega}}{4{\pi}}{\lambda}(x){\delta}^2(0)tr({\gamma})$ and thus a regularization is needed . For example Fujikawa regularization of (\ref{anomalylimit}) was performed in \cite{nagao1} and was shown there to reproduce the correct anomaly . Here however we know from (\ref{ref}) that the action (\ref{anomaly} ) on this finite matrix model is identically zero and this can not be made into anything else which seems in  contradiction with Fujikawa regularization of (\ref{anomalylimit}) .

It is therefore natural to modify appropriately chiral transformations in such a way as to shift the anomaly (\ref{ep}) from the action to the measure . The required deformation is not difficult to find and one obtains
\begin{eqnarray}
&&{\psi}_F{\longrightarrow}{\psi}^{'}_F={\psi}_F+({\Gamma}^{R}{\lambda}{\psi}_F)+O({\lambda}^2)\nonumber\\
&&\bar{\psi}_F{\longrightarrow}\bar{\psi}^{'}_F=\bar{\psi}_F-(\bar{\psi}_F{\lambda}\hat{\Gamma}^L)-(\bar{\psi}_F{\lambda}{\delta}{\Gamma})+O({\lambda}^2),\label{24}
\end{eqnarray}
where the deformation ${\delta}{\Gamma}$ is given by
\begin{eqnarray}
{\delta}{\Gamma}=-\frac{i}{2l+1}{\epsilon}_{abc}{\sigma}_cF_{ab}^F\frac{1}{{\cal D}_{AF}}.
\end{eqnarray}
Remark that (\ref{24}) is also dictated by (\ref{dicta}) which can be put in the equivalent form ${\cal D}_{GF}{\Gamma}^R-(\hat{\Gamma}^L+{\delta}{\Gamma}){\cal D}_{GF}=0$. Remark also that all  needed requirement are satisfied by this deformation : ${\delta}{\Gamma}$ is gauge covariant and drops in the limit and therefore the transformations (\ref{24}) are consistent with gauge symmetry and reduce in the limit to ordinary chiral transformations  . Remark also that we can invert the gauged Dirac operator since we are considering only trivial $U(1)-$bundles over ${\bf S}^2$ , i.e there is no monopole. Under these modified chiral transformations the change in the action is only a total covariant divergence , namely

\begin{eqnarray}
{\Delta}S_{AF}&=&-\frac{1}{2l+1}Tr_{l}{\lambda}[D_a^F,{\cal J}_a^5],
\end{eqnarray}
while the measure becomes non-symmetric , in other words
\begin{eqnarray}
S_{{\theta}F}&=&-\frac{1}{2l+1}Tr_{l}\sum_{\mu}{\phi}^{+}(\mu,A){\lambda}({\Gamma}^R-\hat{\Gamma}^L-{\delta}{\Gamma}){\phi}(\mu,A)\nonumber\\
&=&-\frac{i}{2l+1}{\epsilon}_{abc}({\lambda})^{BC}(F_{ab}^F)^{CD}tr_2{\sigma}_c({\cal D}_{AF}^{-1})^{DA,BA}.\label{alt1}
\end{eqnarray}
This anomaly is of course exactly identical to the result (\ref{alt}) since what we have just done is to shift the anomaly from the action to the measure . Remark also that since ${\Gamma}^R-\hat{\Gamma}^L-{\delta}{\Gamma}{\longrightarrow}-2{\gamma}$ when $l{\longrightarrow}{\infty}$ the continuum limit of the first line of (\ref{alt1}) is still given by (\ref{anomalylimit}) which as we said needs a proper regularization, whereas the second line of (\ref{alt1}) can now be thought of as a regularization of (\ref{anomalylimit}) which is provided in this noncommutative context by the fuzzy sphere .

\paragraph{$3$ Gauge-Invariant Axial Current }
Similarly to the gauge-covariant axial current defined in section $4.3$ we define now gauge-invariant axial current . First we introduce the following chiral transformations
\begin{eqnarray}
&&{\psi}_F{\longrightarrow}{\psi}^{'}_F={\psi}_F+({\Gamma}^{R}{\psi}_F){\lambda}+O({\lambda}^2)\nonumber\\
&&\bar{\psi}_F{\longrightarrow}\bar{\psi}^{'}_F=\bar{\psi}_F-{\lambda}(\bar{\psi}_F\hat{\Gamma}^L)-{\lambda}(\bar{\psi}_F{\delta}{\Gamma})+O({\lambda}^2),\label{25}
\end{eqnarray}
where now the chiral parameter ${\lambda}$ is a matrix in $Mat_{2l+1}$ which does not transform under gauge transformations. Under these gauge-invariant chiral transformations the change in the action is only a total divergence , namely

\begin{eqnarray}
{\Delta}S_{GF}&=&-\frac{1}{2l+1}Tr_{l}{\lambda}[L_a,{J}_a^5],\nonumber\\
J_a^5&=&\bar{\psi}_F{\sigma}_a{\Gamma}^R{\psi}_F.
\end{eqnarray}
The measure is still non-symmetric but its change is now of the form
\begin{eqnarray}
S_{{\theta}F}&=&-\frac{1}{2l+1}Tr_{l}{\lambda}\sum_{\mu}{\phi}^{+}(\mu,A)({\Gamma}^R-\hat{\Gamma}^L-{\delta}{\Gamma}){\phi}(\mu,A)\nonumber\\
&=&-\frac{i}{2l+1}{\epsilon}_{abc}({\lambda})^{BA}(F_{ab}^F)^{CD}tr_2{\sigma}_c({\cal D}_{AF}^{-1})^{DB,CA}.
\end{eqnarray}
The contribution of ${\Gamma}^R-\hat{\Gamma}^L$ vanished by a similar argument to that which led to (\ref{ref}) .
The WT identity ${\Delta}S_{GF}=S_{\theta F}$ will now simply look like
\begin{eqnarray}
[L_a,{J}_a^5]^{AB}=i{\epsilon}_{abc}(F_{ab}^F)^{CD}tr_2{\sigma}_c({\cal D}_{AF}^{-1})^{DB,CA}.
\end{eqnarray}
In this case since $[L_a,J_a^5]^{AB}$ does not transform under gauge transformations , one can immediately conclude that the left hand side must also not transform under gauge transformations and therefore it can only be proportional to the identity , i.e $[L_a,J_a^5]^{AB}{\propto}{\delta}^{AB}$.

\bibliographystyle{unsrt}

\end{document}